\documentclass[twocolumn]{aastex631}
\NewPageAfterKeywords
\usepackage{comment}
\usepackage[flushmargin]{footmisc}

\newcommand{\pokemonnum}{1125}
\newcommand{\observednum}{151}
\newcommand{\percentresolved}{58.9\%}
\newcommand{\percentdetectable}{73.2\%}
\newcommand{\percentdetectablehundred}{70.3\%}

\received{September 22, 2023}
\revised{November 7, 2023}
\accepted{November 10, 2023}

\submitjournal{AJ}

\graphicspath{{./}{figures/}}

\begin{document}

\title{The POKEMON Speckle Survey of Nearby M Dwarfs. II. Observations of \pokemonnum{} Targets}

\correspondingauthor{Catherine A. Clark}
\email{catherine.a.clark@jpl.nasa.gov}

\author[0000-0002-2361-5812]{Catherine A. Clark}
\affil{Jet Propulsion Laboratory, California Institute of Technology, Pasadena, CA 91109 USA}
\affil{NASA Exoplanet Science Institute, IPAC, California Institute of Technology, Pasadena, CA 91125 USA}

\author[0000-0002-8552-158X]{Gerard T. van Belle}
\affil{Lowell Observatory, 1400 West Mars Hill Road, Flagstaff, AZ 86001, USA}

\author[0000-0003-2159-1463]{Elliott P. Horch}
\affil{Southern Connecticut State University, 501 Crescent Street, New Haven, CT 06515, USA}

\author[0000-0003-2527-1598]{Michael B. Lund}
\affil{NASA Exoplanet Science Institute, IPAC, California Institute of Technology, Pasadena, CA 91125 USA}

\author[0000-0002-5741-3047]{David R. Ciardi}
\affil{NASA Exoplanet Science Institute, IPAC, California Institute of Technology, Pasadena, CA 91125 USA}

\author[0000-0002-5823-4630]{Kaspar von Braun}
\affil{Lowell Observatory, 1400 West Mars Hill Road, Flagstaff, AZ 86001, USA}

\author[0000-0001-6031-9513]{Jennifer G.\ Winters}
\affil{Bridgewater State University, 131 Summer St., Bridgewater, MA 02324, USA}
\affil{Center for Astrophysics $\vert$ Harvard \& Smithsonian, 60 Garden Street, Cambridge, MA 02138, USA}

\author[0000-0002-0885-7215]{Mark E. Everett}
\affiliation{NSF’s National Optical-Infrared Astronomy Research Laboratory, 950 N. Cherry Ave., Tucson, AZ 85719, USA}

\author[0000-0003-4236-6927]{Zachary D. Hartman}
\affil{Gemini Observatory/NSF's NOIRLab, 670 A'ohoku Place, Hilo, HI 96720 USA}

\author[0000-0003-4450-0368]{Joe Llama}
\affil{Lowell Observatory, 1400 West Mars Hill Road, Flagstaff, AZ 86001, USA}



\begin{abstract}

Stellar multiplicity is correlated with many stellar properties, yet multiplicity measurements have proven difficult for the M dwarfs -- the most common type of star in our galaxy -- due to their faintness and the fact that a reasonably-complete inventory of later M dwarfs did not exist until recently. We have therefore carried out the Pervasive Overview of ``Kompanions'' of Every M dwarf in Our Neighborhood (POKEMON) survey, which made use of the Differential Speckle Survey Instrument on the 4.3-meter Lowell Discovery Telescope, along with the NN-EXPLORE Exoplanet Stellar Speckle Imager on the 3.5-meter WIYN telescope. The POKEMON sample is volume-limited from M0V through M9V out to 15 pc, with additional brighter targets at larger distances. In total, \pokemonnum{} targets were observed. New discoveries were presented in the first paper in the series. In this second paper in the series, we present all detected companions, gauge our astrometric and photometric precision, and compare our filtered and filterless speckle observations. We find that the majority (\percentresolved{}) of the companions we detect in our speckle images are not resolved in Gaia, demonstrating the need for high-resolution imaging in addition to long-term astrometric monitoring. Additionally, we find that the majority (\percentdetectable{}) of simulated stellar companions would be detectable by our speckle observations. Specifically within 100 au, we find that \percentdetectablehundred{} of simulated companions are recovered. Finally, we discuss future directions of the POKEMON survey.

\end{abstract}



\section{Introduction} \label{sec:introduction}


The smallest, coolest, faintest stars --- the M dwarfs --- are a microcosm of stellar astrophysics. The M-dwarf mass range extends from the hydrogen burning limit of $0.08 M_\odot$ \citep{BaraffeChabrier1996ApJ...461L..51B} at the bottom of the Main Sequence to over half the mass of the Sun -- nearly a factor of eight. Our low-mass neighbors also span nearly a factor of six in radius and over two orders of magnitude in luminosity \citep{PecautMamajek2013ApJS..208....9P}. Additionally, the M dwarfs have lifetimes on the order of the age of the Universe \citep{Laughlin1997ApJ...482..420L}, and remain active and rapidly-rotating for longer than stars of earlier spectral types \citep[e.g.,][]{Popinchalk2021ApJ...916...77P}. Furthermore, the M dwarfs comprise over 70\% of stars in the galaxy \citep{Henry2006AJ....132.2360H, Winters2015AJ....149....5W} and three quarters of the stars within 10 pc \citep{Henry2018AJ....155..265H}, even though not a single one is visible to the naked eye \citep{Shields2016PhR...663....1S}. The M dwarfs are thus critical to our understanding of stellar astrophysics, and in particular, the Solar Neighborhood.

Stars of every spectral type preferentially form in groups \citep[e.g.,][]{Lada1991ApJ...371..171L}. The size of these groups ranges from binary-star systems to clusters containing hundreds of thousands of stars. Multiplicity has been found to be correlated with multiple stellar properties; for example, higher mass stars are more likely to be found in multi-star systems \citep{Mason2009AJ....137.3358M, Raghavan2010ApJS..190....1R, DucheneKraus2013ARA&A..51..269D, Winters2019AJ....157..216W}, and older stars are less likely to be found in multi-star systems \citep{Mason1998AJ....116.2975M}. Binarity can also affect stellar evolution for stars of all masses \citep[e.g.,][]{Iben1991ApJS...76...55I}. Understanding the rate and properties of stellar companions is therefore critical to our understanding of the stars themselves, in particular for the M-dwarfs due to their high occurrence in our galaxy.

Understanding stellar multiplicity is also critical to the detection and characterization of exoplanets. Specifically, determining the occurrence rate of planets is highly-dependent on an understanding of the multiplicity of the host stars \citep{DressingCharbonneau2015ApJ...807...45D}. Known multiplicity is particularly crucial for characterizing exoplanets detected using the transit method. Transit detections from large field-of-view pixel instruments aboard missions such as the Transiting Exoplanet Survey Satellite \citep[TESS;][]{Ricker2015JATIS...1a4003R} can lead to erroneous inferences of the planetary properties if unknown secondary companions are contaminating the light curve. Calculations of planet statistics --and thus our understanding of quantities like $\eta_{\oplus}$ -- therefore hinge critically on the accurate determination of stellar multiplicity \citep{Ciardi2015ApJ...805...16C}.

A number of surveys have been conducted over the past few decades to quantify and understand the rate and properties of the stellar companions that the M dwarfs host \citep[e.g.,][]{Henry1991PhDT........11H, Law2008MNRAS.384..150L, Janson2012ApJ...754...44J, Jodar2013MNRAS.429..859J, Ward-Duong2015MNRAS.449.2618W, Winters2019AJ....157..216W, Cifuentes2023PhDT.........1C}. However, many of these studies were limited by the size of their sample, due to the fact that the M dwarfs are quite faint in the optical, and that a reasonably-complete inventory of later M dwarfs did not even exist until recently \citep{Kirkpatrick2014ApJ...783..122K, LuhmanSheppard2014ApJ...787..126L, Winters2021AJ....161...63W}. Additionally, many of these studies were not sensitive to close-in stellar companions due to the resolution limits of the surveys. Close-in companions are known to affect planet formation mechanisms \citep{HaghighipourRaymond2007ApJ...666..436H, Jang-Condell2015ApJ...799..147J, RafikovSilsbee2015ApJ...798...69R, RafikovSilsbee2015ApJ...798...70R}, and it has been shown that both solar-type \citep[e.g.,][]{Howell2021AJ....161..164H} and M-type \citep{Clark2022AJ....163..232C} exoplanet host stars have fewer close-in stellar companions than their non-planet-hosting counterparts. A complete census of M-dwarf companions at all separations is therefore necessary to determine statistics on their properties and occurrence rates. This presents an opportunity to use high-resolution imaging to survey a large number of M dwarfs and their companions at separations down to the diffraction limit.

We have therefore carried out the Pervasive Overview of ``Kompanions'' of Every M dwarf in Our Neighborhood (POKEMON) speckle survey of nearby M dwarfs. The POKEMON survey is volume-limited out to 15 pc from M0V through M9V, with additional brighter targets at larger distances. In total, we have imaged \pokemonnum{} M-dwarfs at diffraction-limited resolution, and we have detected companions to \observednum{} of these targets. The first paper in the series \citep{Clark2022AJ....164...33C} presented the new discoveries, and this second paper in the series presents all detections from the POKEMON survey.

In Section \ref{sec:observations}, we describe how we defined the POKEMON sample, our observational routine, and our data reduction process. In Section \ref{sec:results}, we present the stellar companions detected throughout the POKEMON survey and examine our astrometric and photometric precision. We also compare our filtered and filterless speckle observations. In Section \ref{sec:discussion}, we evaluate the effect of stellar multiplicity on Gaia astrometry and assess any potentially-missed companions. Finally, in Section \ref{sec:conclusions}, we summarize our conclusions and detail the future work to be done on the POKEMON survey.

\clearpage

\section{Observations} \label{sec:observations}

In this section we describe how we defined the POKEMON sample. We also characterize our observational routine and our data reduction process.

\subsection{Definition of the Sample}

The underlying, essential goal of our target selection process was to provide a complete-as-possible sample of M dwarfs out to 15 pc, with special care taken to include the previously-overlooked later-type M dwarfs. Assuming a homogeneous distribution of stars in the Solar Neighborhood, targets from Gaia or other catalogs can be reviewed to investigate whether their number increases linearly with volume, or with the cube of distance ($d$). This is known as a $d^3$ trend. If a catalog is volume-complete, then the $d^3$ trend should remain constant with increasing volume; when a catalog overlooks faint targets, there are deviations from this trend. An examination of the Catalog of Nearby Stars \citep[CNS3;][]{GlieseJahreiss1991adc..rept.....G} indicated a possible deviation from the targets per unit volume $d^3$ line at 15 pc, with a definite fall-off trend (i.e. incompleteness) at 25 pc. A number of newer catalogs were therefore cross-referenced to eliminate as much incompleteness as possible. This included nearby neighbor discoveries from RECONS \citep{Henry2006AJ....132.2360H, Winters2015AJ....149....5W} and the Two Micron All Sky Survey \citep[2MASS;][]{Skrutskie2006AJ....131.1163S}. Of particular relevance were the results from the AllWISE Motion Survey as reported by \citet{Kirkpatrick2014ApJ...783..122K} and \citet{LuhmanSheppard2014ApJ...787..126L}, which had found 3,525 and 762 high-proper-motion objects, respectively. Using the spectral type-color relationships from \citet{LuhmanSheppard2014ApJ...787..126L}, we identified $\sim350$ additional potential objects with colors indicating spectral types between M4V and M9V. Additional surveys that were cross-referenced include the Database of Ultracool Parallaxes based on the Hawaii Infrared Parallax Program \citep{DupuyLiu2012ApJS..201...19D}, the CARMENES input catalog \citep{Alonso-Floriano2015AA...577A.128A}, the APOGEE input catalog \citep{Deshpande2013AJ....146..156D}, and the 2016 release of the Pan-STARRS Parallax and Proper Motion Catalog \citep{Waters2015IAUGA..2256019W}, which had the specific focus of identifying nearby low-mass stars \citep{Magnier2015IAUGA..2257922M}.

\begin{figure*}
    \centering
    \includegraphics[width=\textwidth]{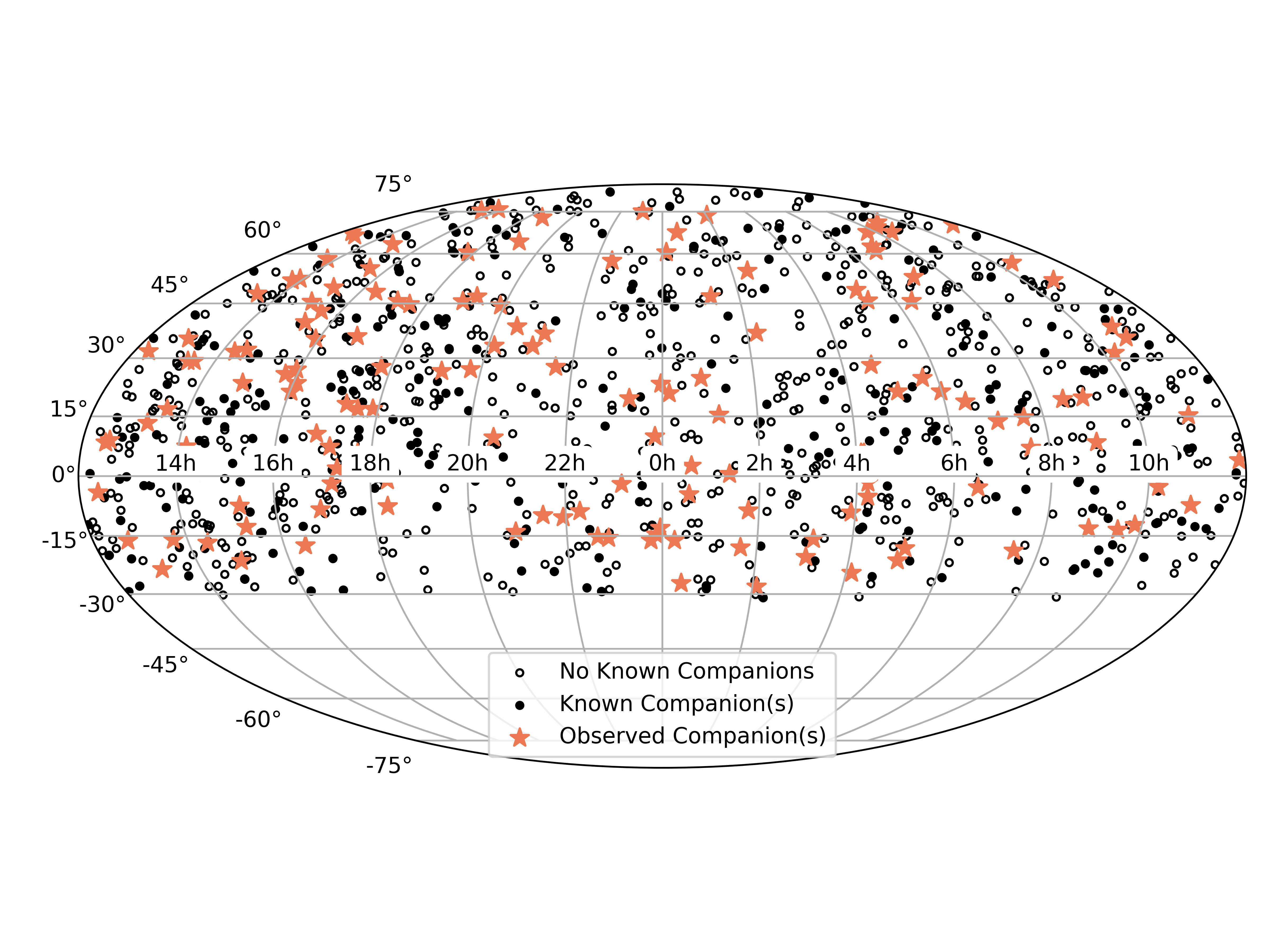}
    \caption{The sky locations of the \pokemonnum{} targets in the POKEMON sample. Targets with no known companions are marked with open black circles, targets with known companions (but without companions detected by us) are marked with filled black circles, and targets with companions detected by us are marked with larger, orange stars.}
    \label{fig:aitoff}
\end{figure*}

At that time, the speckle cameras available to the POKEMON survey were located at northern hemisphere facilities, necessitating a declination cut of $\delta > -30^{\circ}$. A cut at $I<15.5$ was also necessary for most objects due to the faint limit of the instruments. In total, our sample consists of \pokemonnum{} M dwarfs north of $\delta > -30^{\circ}$, brighter than $I=15.5$ mag, and within $\sim100$ pc (Figure \ref{fig:aitoff}). We note that the sample size is updated from the previous paper in the series.

We characterize the \pokemonnum{} stars in the POKEMON sample in Table \ref{table:targets}, where we include the 2MASS ID or name, Gaia Data Release 3 \citep[DR3;][]{Gaia2023AA...674A...1G} ID, Gaia $G$ magnitude, Gaia $G_{RP}$ magnitude, parallax-derived distance, reference for the distance, Gaia re-normalized unit weight error (RUWE) value, and the Gaia ipd\_frac\_multi\_peak (IPDFMP) value. We also note whether the target had companions detected by us.

\begin{figure*}
    \centering
    \includegraphics[width=0.49\textwidth]{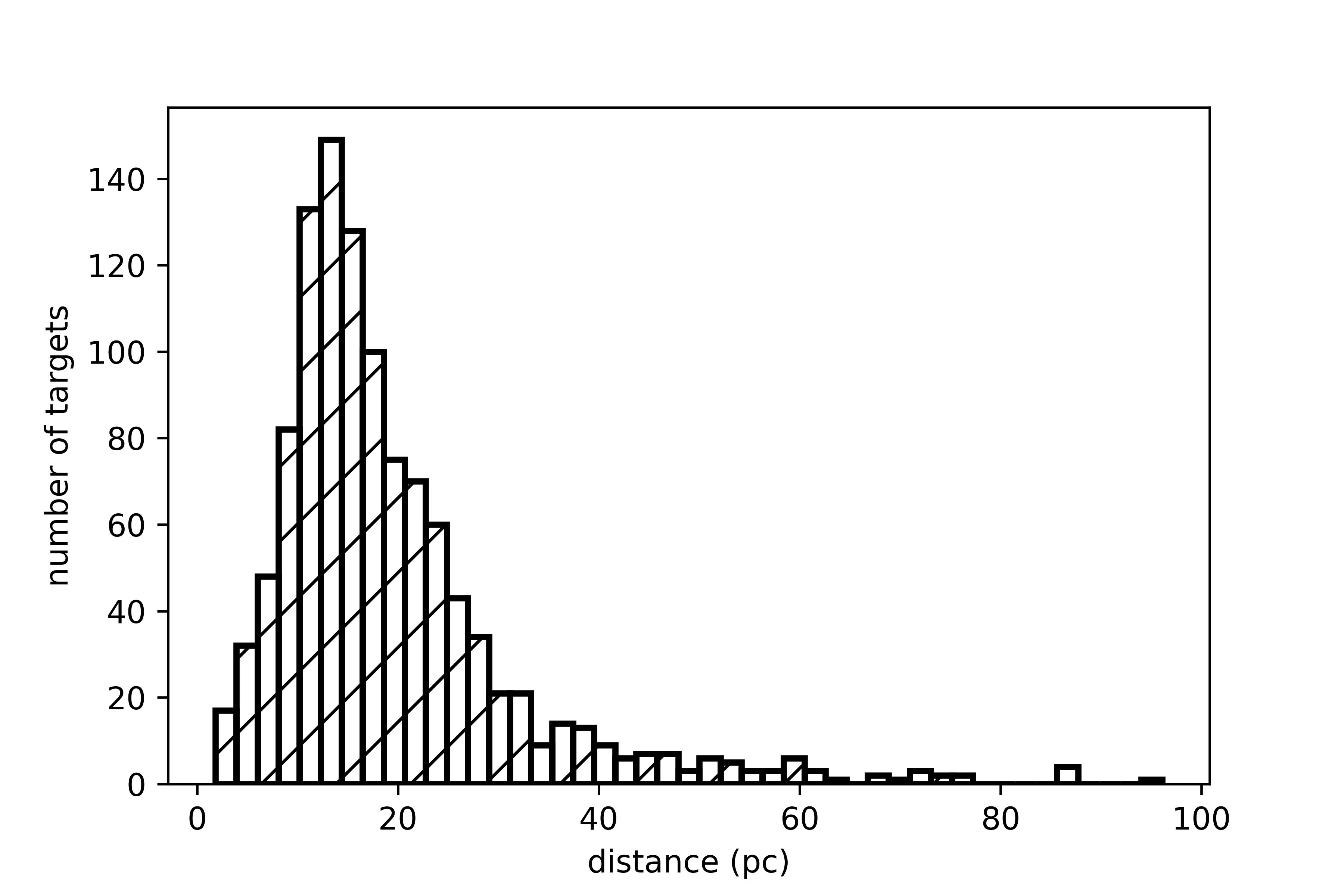}
    \includegraphics[width=0.49\textwidth]{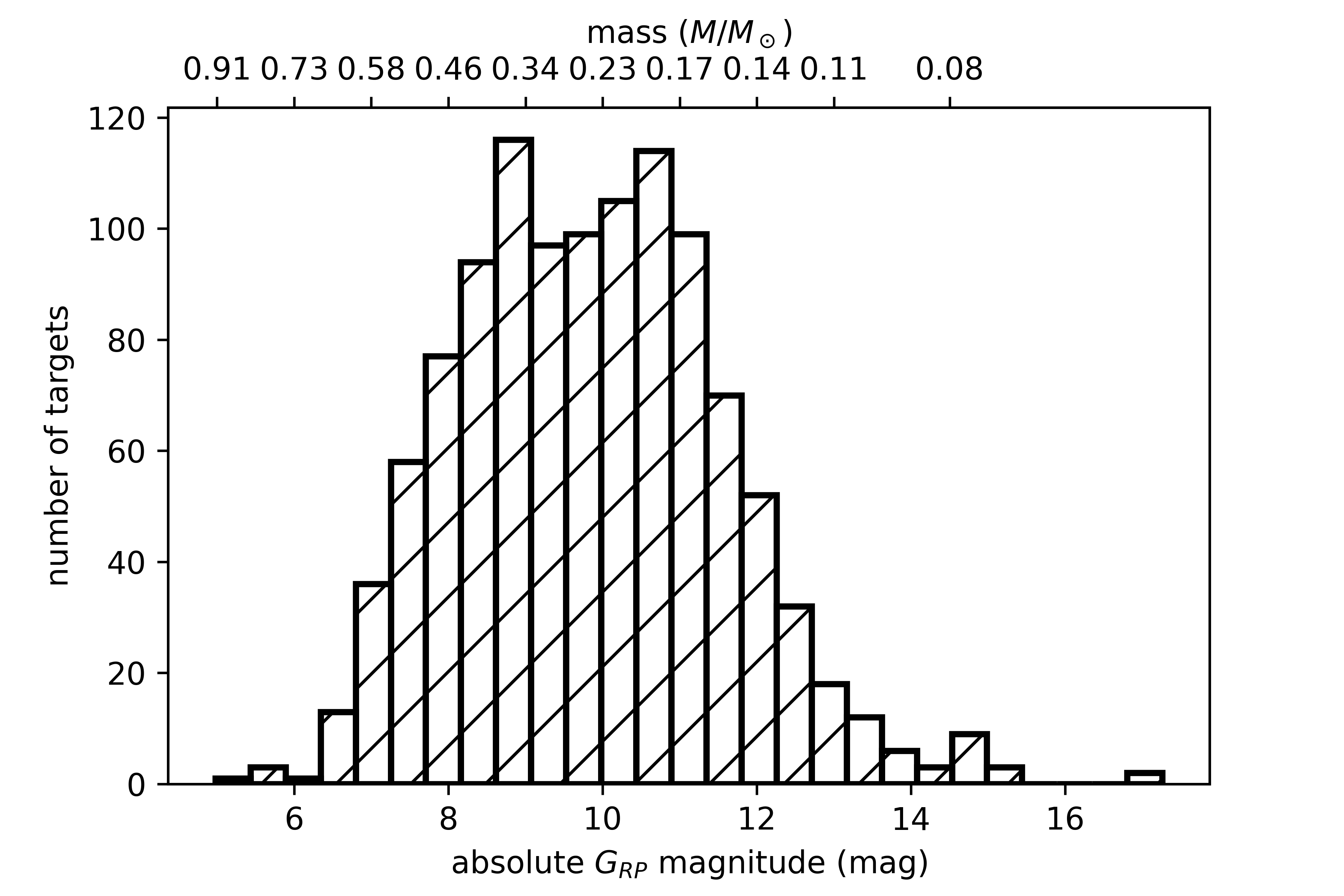}
    \caption{Distance (left) and absolute $G_{RP}$ magnitude distributions for the \pokemonnum{} targets in the POKEMON sample. Using the mass-magnitude relation from \citet{GiovinazziBlake2022AJ....164..164G}, we also show the estimated masses for these targets.}
    \label{fig:histograms}
\end{figure*}

In Figure \ref{fig:histograms} we show histograms of the distances and absolute $G_{RP}$ magnitudes for the targets in the POKEMON sample. We use the absolute $G_{RP}$ magnitudes and the mass-magnitude relation from \citet{GiovinazziBlake2022AJ....164..164G} to estimate masses for the POKEMON targets.

Using the apparent $G$ magnitudes from Gaia, the stellar distances, and the reference stellar properties from \citet{PecautMamajek2013ApJS..208....9P}, we estimate spectral types for the POKEMON targets in Table \ref{table:spectral_types}. We note that some of the POKEMON targets are estimated to be either K dwarfs or L dwarfs; the fourth paper in the series will provide homogenous spectral types for all POKEMON targets, allowing for more accurate stellar characterization, and establishing the M-dwarf multiplicity rate by spectral subtype through M9V for the first time.

\begin{deluxetable*}{ccccccccc}
\tablecaption{POKEMON targets
\label{table:targets}}
\tablehead{\colhead{2MASS ID or Name} & \colhead{Gaia DR3 ID} & \colhead{$G$ Magnitude} & \colhead{$G_{RP}$ Magnitude} & \colhead{Distance} & \colhead{Reference} & \colhead{RUWE} & \colhead{IPDFMP} & \colhead{Companion(s) detected?} \\ 
\colhead{} & \colhead{} & \colhead{(mag)} & \colhead{(mag)} & \colhead{(pc)} & \colhead{} & \colhead{} & \colhead{} & \colhead{}}
\rotate
\startdata
00051079+4547116 & 386655019234959872 & 9.1 & 8.0 & 11.5 & 8 & 1.0 & 0 & N \\
00054090+4548374 & 386653747925624576 & 8.3 & 7.4 & 11.5 & 8 & 1.2 & 2 & N \\
00064325-0732147 & 2441630500517079808 & 11.8 & 10.4 & 4.8 & 8 & 1.3 & 4 & N \\
00074264+6022543 & 429297924157113856 & 12.4 & 10.7 & 15.4 & 8 & 3.9 & 72 & Y \\
00085391+2050252 & 2798766647610195456 & 12.0 & 10.7 & 18.1 & 7 &  & 8 & Y \\
00085512+4918561 & 393621524910343296 & 14.4 & 13.0 & 14.8 & 8 & 1.1 & 0 & N \\
00113182+5908400 & 423027104407576576 & 13.5 & 12.1 & 9.3 & 8 & 1.1 & 0 & N \\
00152799-1608008 & 2368293487261055488 & 10.3 & 8.9 & 5.0 & 16 &  & 60 & Y \\
00153905+4735220 & 392350008432819328 & 10.0 & 8.9 & 11.5 & 8 & 0.9 & 3 & N \\
00154919+1333218 & 2768048564768256512 & 11.4 & 10.2 & 12.2 & 8 & 1.2 & 0 & N \\
\enddata
\tablecomments{Table \ref{table:targets} is published in its entirety in the machine-readable format. A portion is shown here for guidance regarding its form and content.}
\tablerefs{(1) \citet{Cifuentes2020AA...642A.115C}; (2) \citet{Dittmann2014ApJ...784..156D}; (3) \citet{DupuyLiu2017ApJS..231...15D}; (4) \citet{Dupuy2019AJ....158..174D}; (5) \citet{FinchZacharias2016AJ....151..160F}; (6) \citet{Finch2018AJ....155..176F}; (7) \citet{Gaia2018AA...616A...1G}; (8) \citet{Gaia2023AA...674A...1G}; (9) \citet{GlieseJahreiss1991adc..rept.....G}; (10) \citet{Henry2006AJ....132.2360H}; (11) \citet{Lepine2009AJ....137.4109L}; (12)  \citet{Murray1986MNRAS.223..629M}; (13) \citet{Riedel2010AJ....140..897R}; (14) \citet{Torres2010AARv..18...67T}; (15) \citet{vanAltena1995gcts.book.....V}; (16) \citet{vanLeeuwen2007AA...474..653V}; (17) \citet{Weinberger2016AJ....152...24W}; (18) \citet{Winters2015AJ....149....5W}; (19) \citet{Winters2017AJ....153...14W}.}
\end{deluxetable*}

\clearpage

\startlongtable
\begin{deluxetable}{cc}
\tablecaption{Spectral types of the targets in the POKEMON sample
\label{table:spectral_types}}
\tablehead{\colhead{Spectral Type} & \colhead{Number of Targets}}
\startdata
K & 52 \\
M0 & 69 \\
M1 & 77 \\
M2 & 129 \\
M3 & 296 \\
M4 & 257 \\
M5 & 203 \\
M6 & 7 \\
M7 & 18 \\
M8 & 6 \\
M9 & 1 \\
L & 8 \\
\enddata
\end{deluxetable}

\subsection{Observational Routine}

We imaged the \pokemonnum{} M dwarfs in the POKEMON sample over 50 nights between UT 2017 April 7 and UT 2020 February 10. These observations were mainly carried out using the Differential Speckle Survey Instrument \citep[DSSI;][]{Horch2009AJ....137.5057H} on the 4.3-meter Lowell Discovery Telescope  \citep[LDT;][]{Levine2022SPIE12182E..27L} located in Happy Jack, AZ. We also used the NN-EXPLORE Exoplanet Stellar Speckle Imager \citep[NESSI;][]{Scott2018PASP..130e4502S} on the 3.5-meter WIYN telescope\footnote{The WIYN Observatory is a joint facility of the NSF's National Optical-Infrared Astronomy Research Laboratory, Indiana University, the University of Wisconsin-Madison, Pennsylvania State University, the University of Missouri, the University of California-Irvine, and Purdue University.} at Kitt Peak National Observatory located outside Tucson, AZ. In Table \ref{table:observing_runs} we list the dates of the observing runs and how many targets were observed during each run. We note that the POKEMON targets were not the only stars observed during each run; with the high cadence of speckle observations and the superior pointing of the LDT and WIYN, a hundred targets or more can easily be observed per night.

\clearpage

\startlongtable
\begin{deluxetable*}{cccc}
\tablecaption{Observing runs
\label{table:observing_runs}}
\tablehead{\colhead{First Night of Observing Run} & \colhead{Final Night of Observing Run} & \colhead{Instrument} & \colhead{Targets Observed} \\
\colhead{(UT Date)} & \colhead{(UT Date)} & \colhead{} & \colhead{}}
\startdata
2017 April 7 & 2017 April 7 & DSSI \tablenotemark{*} & 29 \\
2017 April 9 & 2017 April 17 & DSSI \tablenotemark{*} & 508 \\
2017 May 4 & 2017 May 8 & DSSI & 87 \\
2017 October 18 & 2017 October 18 & DSSI & 50 \\
2017 October 20 & 2017 October 21 & DSSI & 127 \\
2018 January 28 & 2018 February 1 & DSSI & 219 \\
2018 August 2 & 2018 August 5 & NESSI & 31 \\
2018 August 26 & 2018 August 26 & NESSI & 6 \\
2018 August 26 & 2018 August 27 & DSSI & 32 \\
2018 August 28 & 2018 August 29 & NESSI & 28 \\
2018 November 17 & 2018 November 17 & NESSI & 3 \\
2018 November 19 & 2018 November 22 & NESSI & 22 \\
2019 January 18 & 2019 January 20 & NESSI & 15 \\
2019 January 22 & 2019 January 25 & NESSI & 22 \\
2019 September 10 & 2019 September 10 & DSSI & 9 \\
2019 September 13 & 2019 September 14 & DSSI & 75 \\
2019 September 16 & 2019 September 16 & DSSI & 2 \\
2020 February 9 & 2020 February 10 & DSSI & 16 \\
\enddata
\tablenotetext{*}{ Filterless observations}
\end{deluxetable*}

DSSI and NESSI produce diffraction-limited images from speckle patterns observed simultaneously at two wavelengths. Each instrument uses a dichroic filter to split the collimated beam from the telescope at $\sim700$ nm into two channels that are imaged on separate high-speed readout Electron-Multiplying Charge Coupled Devices. DSSI uses two narrow-band filters centered at 692 and 880 nm, with filter widths of 40 and 50 nm, respectively. NESSI uses filters centered at 562 and 832 nm, with filter widths of 44 and 40 nm, respectively.

The raw data from DSSI and NESSI are image cubes consisting of a sequence of 1000 40-millisecond exposures. These short exposures are necessary to ``freeze'' out the atmospheric interference from the observations, and to obtain high contrast in the speckles. Bright objects ($V<11$) require only $\sim1-2$ minutes of observing time, during which one data cube is acquired, while fainter targets require up to $\sim10$ minutes of observing time, during which up to nine data cubes are acquired. Standard observing also includes periodic observations of bright, unresolved, single stars from the Bright Star Catalog \citep{HoffleitJaschek1982bsc..book.....H} to probe the atmospheric conditions experienced by the target of interest. All data cubes are stored as multi-extension FITS files.

The pixel scale and image orientation are empirically confirmed by observing binaries with extremely well-known orbits (those listed as Grade 1 in the Sixth Orbit Catalog; \citealt{Hartkopf2001AJ....122.3472H}). Their ephemeris positions are computed based on the orbital elements, and their scale and orientation are derived from these results. For more information on calibration and overall speckle data quality obtainable at the LDT and WIYN, see \citet{Horch2021AJ....161..295H} and earlier papers in that series.

Each target was observed at least once, though some targets were observed twice, or even three or four times.

\subsection{Data Reduction} \label{subsec:data_reduction}

The data were reduced with an updated version of the pipeline described in \citet{Horch2009AJ....137.5057H, Horch2011AJ....141...45H, Horch2011AJ....141..180H}, which uses bispectral analysis \citep{Lohmann1983ApOpt..22.4028L} to compute a reconstructed image from the data cube(s) collected for each target.

Multi-star systems produce fringe patterns in the Fourier plane, and we use this fact to probe the properties of each star in the system. A two-dimensional autocorrelation function is calculated for each 40-millisecond exposure in the data cube, and summed over all exposures. The Fourier transform of the autocorrelation function is calculated and squared in order to obtain the power spectrum, which is then normalized. After dividing by the power spectrum of a point source, the residual two-dimensional power spectrum appears as a set of fringes for each pair of stars in the field of view. The fringes are fit using a cosine-squared function to determine the relative astrometry and photometry of any pairs. The reconstructed image is constructed from the object's modulus, and the phase estimate is obtained from the bispectrum.

We note that the reconstructed images generated from our speckle observations are limited. They often contain three peaks, which derive from low signal-to-noise in the phase portion of the calculation. For our reconstructed images, we start with a ``zero-phase'' assumption and relax to the final phase map \citep{Meng1990JOSAA...7.1243M}. When there is low signal-to-noise in the phase, the zero-phase assumption results in a phase map that is close to zero across the Fourier plane, resulting in a lack of phase information. The reconstructed image therefore appears similar to an inverse Fourier transform of the square root of the power spectrum. Therefore for this work, we mainly use the reconstructed images for identifying whether a star is single or multiple.

By examining annuli in the reconstructed image that are centered on the primary star, all local maxima and minima in the annulus are determined, and their mean values and standard deviations are derived. The detection limit within each annulus is estimated as the mean value of the maxima plus five times the average sigma of the maxima and minima. A ``contrast curve'' is then produced by calculating the detection limit within each annulus as a function of separation \citep[e.g., Figure 6 in][]{Horch2011AJ....141...45H}.

We note that the contrast curves are rough estimates that can contain systematic error, particularly in the low signal-to-noise regime as described previously. Therefore for this work, we mainly use the contrast curves to place limits on the types of companions that could still exist around ``single'' stars.

\begin{figure*}
    \centering
    \includegraphics[width=0.49\textwidth]{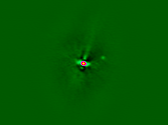}
    \includegraphics[width=0.49\textwidth]{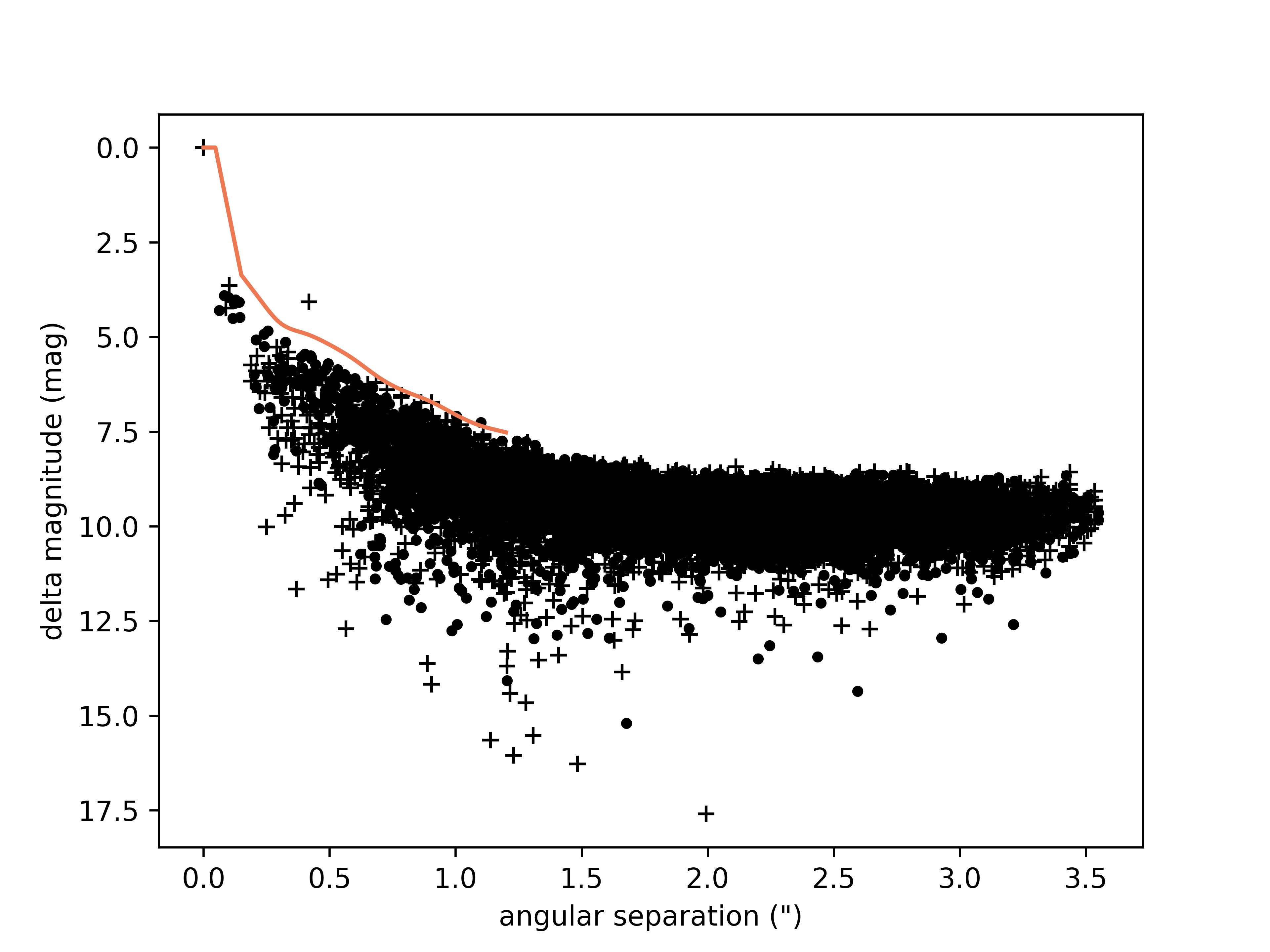}
    \includegraphics[width=0.49\textwidth]{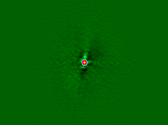}
    \includegraphics[width=0.49\textwidth]{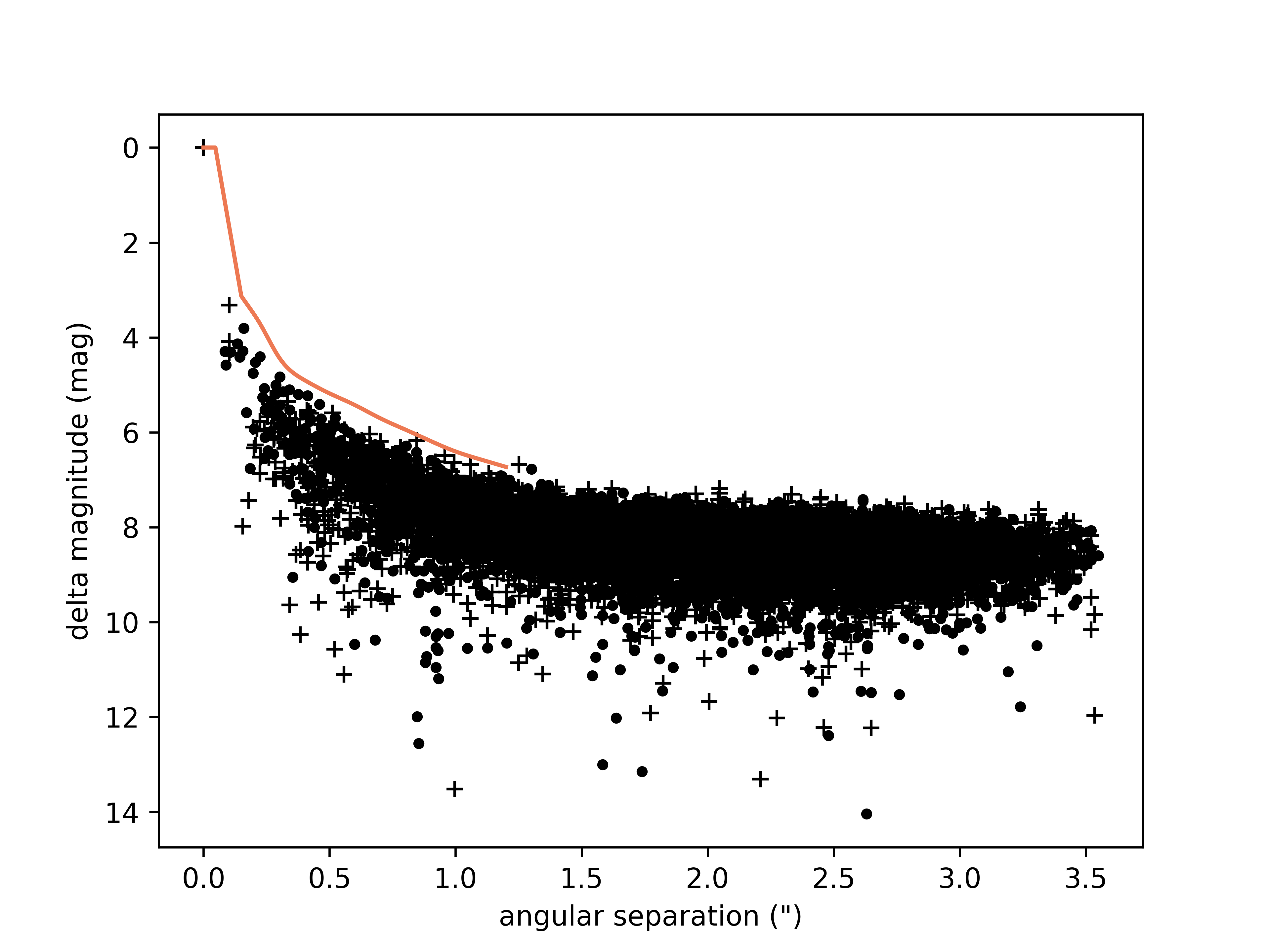}
    \caption{\textit{Top:} The reconstructed image and contrast curve for 2MASS J08313759+1923395 and its companion at a separation of $0.3397\arcsec$ that is fainter by 5.21 magnitudes at $\lambda>700$ nm. This is the largest delta magnitude we measured that was not an upper limit. The companion is marked by a $+$ sign above the contrast curve. \textit{Bottom:} The reconstructed image and contrast curve for 2MASS J08294949+2646348 at $\lambda>700$ nm. We have determined that this star is single based on the contrast limits derived from the reconstructed image.}
    \label{fig:RIs}
\end{figure*}

\begin{figure*}
    \centering
    \includegraphics[width=\textwidth]{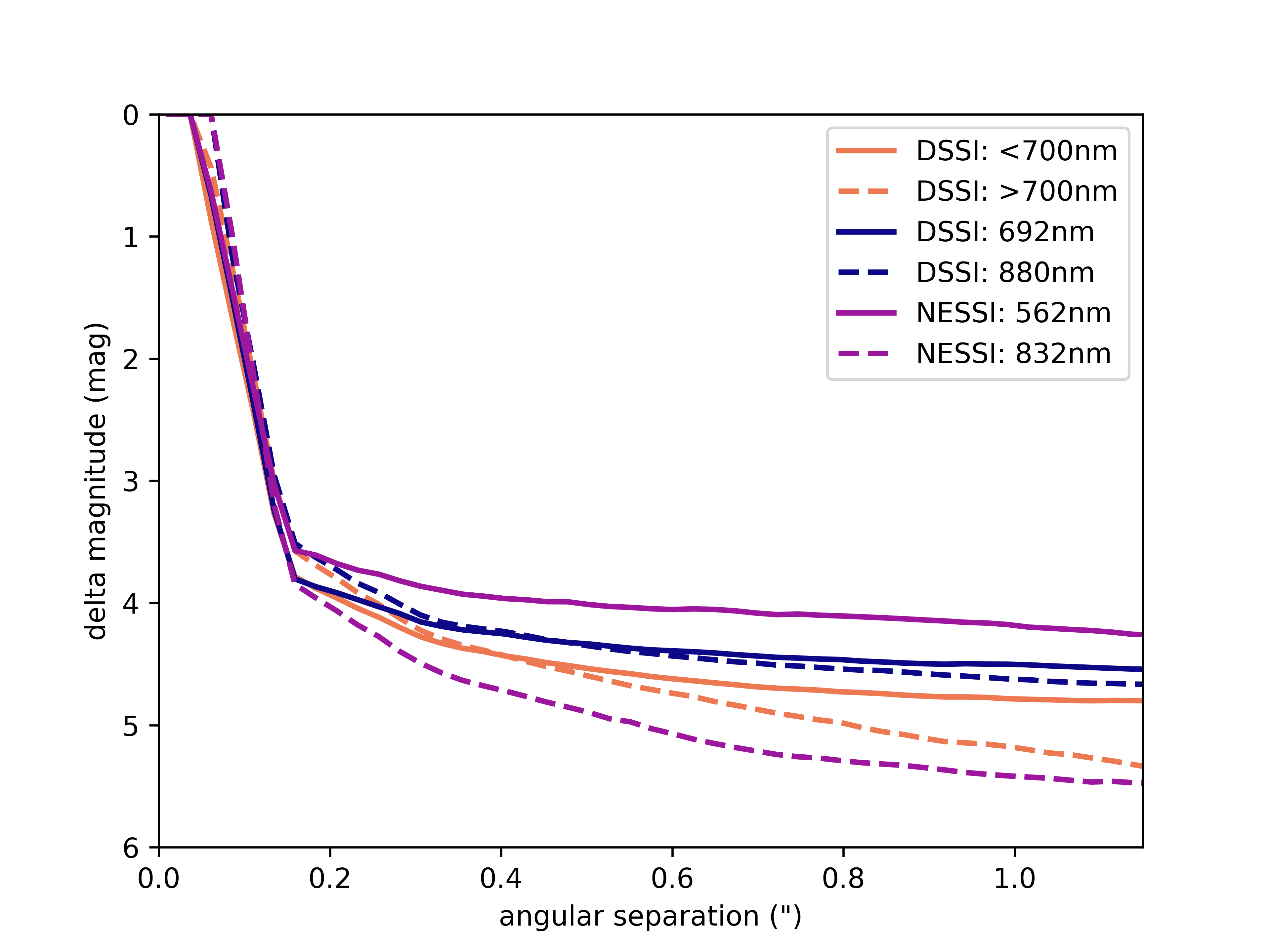}
    \caption{Median contrast curves for the $<700$, $>700$, 692, and 880 nm DSSI observations and the 562 and 832 nm NESSI observations taken throughout the POKEMON survey.}
    \label{fig:median_contrast_curves}
\end{figure*}

Two example reconstructed images and their corresponding contrast curves are shown in Figure \ref{fig:RIs}. The median contrast curves from all DSSI and NESSI observations are shown in Figure \ref{fig:median_contrast_curves}. These contrast curves show that we typically achieve a spatial resolution of $\sim40$ mas, which is comparable to the resolution of the near-infrared adaptive optics system on the 10-meter Keck II Telescope. Because of our achieved spatial resolution and the proximity of the POKEMON targets, we were able to identify stellar companions in the reconstructed images as close-in as $\sim1$ au.

\clearpage

\section{Results} \label{sec:results}

In this section we present the companions detected throughout the POKEMON survey, as well as our evaluation of our astrometric and photometric precision.

\subsection{Detected Companions}

In this section we report the observed properties for the \observednum{} M dwarfs that had a companion detected throughout the POKEMON survey, including those from \citet{Clark2022AJ....164...33C}, \citet{Clark2022RNAAS...6..242C}, and \citet{Clark2023RNAAS...7..206C}. We will present companions that are known to the literature but were not detected in our speckle images in the third paper in the series (Clark et al. submitted). Nonetheless, we have cross-matched the POKEMON sample with the Washington Double Star \citep[WDS;][]{Mason2001AJ....122.3466M} catalog to guide the reader to previous epochs of observation for these multiples. Additionally, this cross-match allowed us to compare our astrometry and photometry with published values to ensure that the motion is consistent with a bound pair; this was the case for all multiples.

The properties of the detected companions are recorded in Table \ref{table:detected_companions}, where we have included the 2MASS ID or name of the primary, WDS ID, epoch of observation (measured in Besselian years), bandpass central wavelength ($\lambda$), bandpass width ($\Delta\lambda$), position angle ($\theta$), angular separation ($\rho$), and magnitude difference ($\Delta m$). We also list the position angle, angular separation, and magnitude difference for the second companion if the system is trinary.

We note several considerations about the content of Table \ref{table:detected_companions} and updates from the previous paper in the series \citep{Clark2022AJ....164...33C}:
\begin{itemize}
    \item In some cases, in particular for the fainter objects, there may be a 180$^{\circ}$ ambiguity in the listed position angles due to the limitations of the reconstructed images described in Section \ref{subsec:data_reduction}.
    \item As shown in previous LDT speckle papers \citep{Horch2015AJ....150..151H, Horch2020AJ....159..233H}, when the seeing value multiplied by the angular separation is larger than 0.6 arcseconds squared, then there may be a systematic error in the photometry, and the observed magnitude difference should be considered as an upper limit; i.e., the companions may be brighter than the quoted delta magnitude value. This is denoted by a $<$ limit flag on the delta magnitude.
    \item As discussed in Section \ref{subsec:filterless_observations}, during the April 2017 observing runs, the narrow-band speckle filters were not installed within DSSI. For these ``filterless'' observations, rather than listing the bandpass and bandpass width of the observation, we instead indicate whether the companion was detected in the $\lambda<700$ nm image or the $\lambda>700$ nm image. We are using these identifiers rather than the ``blue'' and ``red'' identifiers used in the previous paper in the series to avoid confusion with the UBVRI photometric system.
    \item We introduce updated properties for several detections presented in the previous paper in the series.
    \item In the previous paper in the series we mistakenly attributed the discovery of several companions to our own work; we note here the correct references for discoveries of these companions: 2MASS J07011725+1348085 \citep{El-Badry2021MNRAS.506.2269E}, 2MASS J11030845+1517518 \citep{Bowler2019ApJ...877...60B}, 2MASS J12435889-1614351 \citep{Vrijmoet2022AJ....163..178V}, 2MASS J15471513+0149218 \citep{Salama2021AJ....162..102S}, and 2MASS J21011610+3314328 \citep{Cortes-Contreras2017AA...597A..47C}.
    \item We include our observations of the companions to 2MASS J10494561+3532515 and 2MASS J17335314+1655129, but we note that the motion of the companions appears linear rather than orbital, indicating that the companions are likely unbound. This was determined using multiple published observations of 2MASS J10494561+3532515, and from unpublished adaptive optics observations of 2MASS J17335314+1655129 (private communication, C. Gelino and J. D. Kirkpatrick). We do not include these companions in the analysis described in Section \ref{sec:discussion}.
    \item As noted in the previous paper in the series, the companion we detected to the late M dwarf 2MASS J13092185-2330350 is likely a brown dwarf, so this companion is not included in the analysis described in Section \ref{sec:discussion}.
    \item We later found 2MASS J14235017-1646116 to be in a common proper motion pair with a more-massive M dwarf, so the new companion we reported in the previous paper in the series makes the system triple.
    \item We have imaged and obtained properties for the companion to 2MASS J15411642+7559347 for the first time.
\end{itemize}

\startlongtable
\begin{deluxetable*}{ccccccccccc}
\tablecaption{Properties of companions detected throughout the POKEMON survey
\label{table:detected_companions}}
\tablehead{\colhead{2MASS ID or name} & \colhead{WDS ID} & \colhead{Epoch} & \colhead{$\lambda$} & \colhead{$\Delta\lambda$} & \colhead{$\theta_1$} & \colhead{$\rho_1$} & \colhead{$\Delta m_1$} & \colhead{$\theta_2$} & \colhead{$\rho_2$} & \colhead{$\Delta m_2$} \\ 
\colhead{} & \colhead{} & \colhead{(2000+)} & \colhead{(nm)} & \colhead{(nm)} &\colhead{($^{\circ}$)} & \colhead{($\arcsec$)} & \colhead{(mag)} & \colhead{($^{\circ}$)} & \colhead{($\arcsec$)} & \colhead{(mag)}}
\startdata
00074264+6022543 & 00077+6022 & 18.5887 & 562 & 44 & 107.6 & 0.9750 & 0.95 &  &  &  \\
 & & 18.5887 & 832 & 40 & 107.9 & 0.9774 & 0.76 &  &  &  \\
00085391+2050252 & 00089+2050 & 18.5853 & 832 & 40 & 232.4 & 0.1329 & 0.64 &  &  &  \\
00152799-1608008 & 00155-1608 & 17.8025 & 692 & 40 & 131.6 & 0.1944 & 2.46 &  &  &  \\
 & & 17.8025 & 880 & 50 & 131.1 & 0.1910 & 1.64 &  &  &  \\
 & & 18.6571 & 832 & 40 & 4.8 & 0.2146 & 1.58 &  &  &  \\
00244419-2708242 & 00247-2653 & 18.6571 & 832 & 40 & 51.7 & 0.9122 & $<2.41$ & & &  \\
00322970+6714080 & 00321+6715 & 18.5887 & 562 & 44 & 354.7 & 0.4695 & 3.94 & 185.4 & 3.5795 & $<2.78$ \\
 & & 18.5887 & 832 & 40 & 355.3 & 0.4806 & 3.05 & 185.6 & 3.5909 & $<2.46$ \\
 & & 19.0643 & 832 & 40 & 3.8 & 0.4941 & 2.44 &  &  &  \\
\enddata
\tablecomments{Table \ref{table:detected_companions} is published in its entirety in the machine-readable format. A portion is shown here for guidance regarding its form and content.}
\tablecomments{Astrometric and photometric uncertainties are described in Sections \ref{subsec:astrometric_precision} and \ref{subsec:photometric_precision}, respectively.}
\end{deluxetable*}

The minimum angular separation we measured was $0.0787\arcsec$, and the maximum angular separation we measured was $3.5909\arcsec$. The minimum delta magnitude we measured was 0 mag (at $\lambda=880$ nm), and the maximum delta magnitude we measured that was not an upper limit was 5.21 (at $\lambda>700$ nm). These wide ranges demonstrate the power and utility of speckle imagers to detect stellar companions, even on mid-sized (3.5- and 4.3-meter) telescopes.

\subsection{Astrometric Precision} \label{subsec:astrometric_precision}

To characterize our astrometric precision, we use the fact that DSSI and NESSI observe simultaneously at two wavelengths. This allow us to calculate the residuals in our astrometry by computing the difference between the observed angular separation and position angle in each channel. The angular separation residuals have an average value of 3.1 mas, with a standard deviation of 12 mas. The position angle residuals have an average value of 0.10$^{\circ}$, with a standard deviation of 1.1$^{\circ}$. The residuals result from the subtraction of two independent measurements with presumably the same uncertainty, so the subtraction has an uncertainty that is $\sqrt{2}$ larger than the uncertainty of either individual measure. This means that the uncertainty in a single angular separation measure is given by 12 divided by $\sqrt{2}$, or 8.4 mas. The average uncertainty in position angle is then 0.77$^{\circ}$. These values are larger than those derived in \citet{Horch2017AJ....153..212H}, \citet{Colton2021AJ....161...21C}, or \citet{Horch2021AJ....161..295H}, which also used DSSI and NESSI; this is likely due to the faintless of our targets, and the filterless observations that took place during the 2017 April observing run. In general, filterless observations reduce the precision of our astrometry, but allow us to observe fainter companions.

\begin{figure*}
    \centering
    \includegraphics[width=0.49\textwidth]{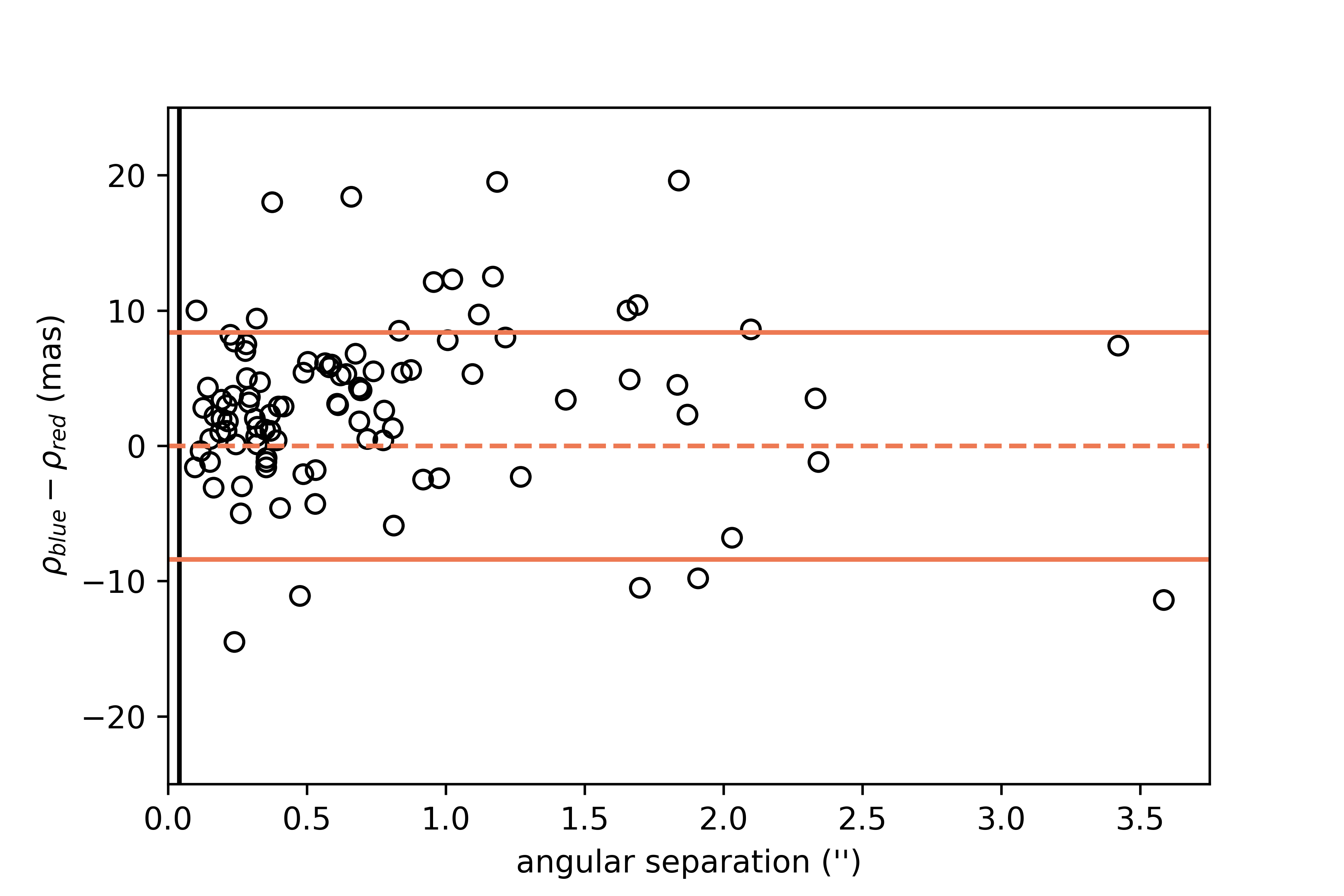}
    \includegraphics[width=0.49\textwidth]{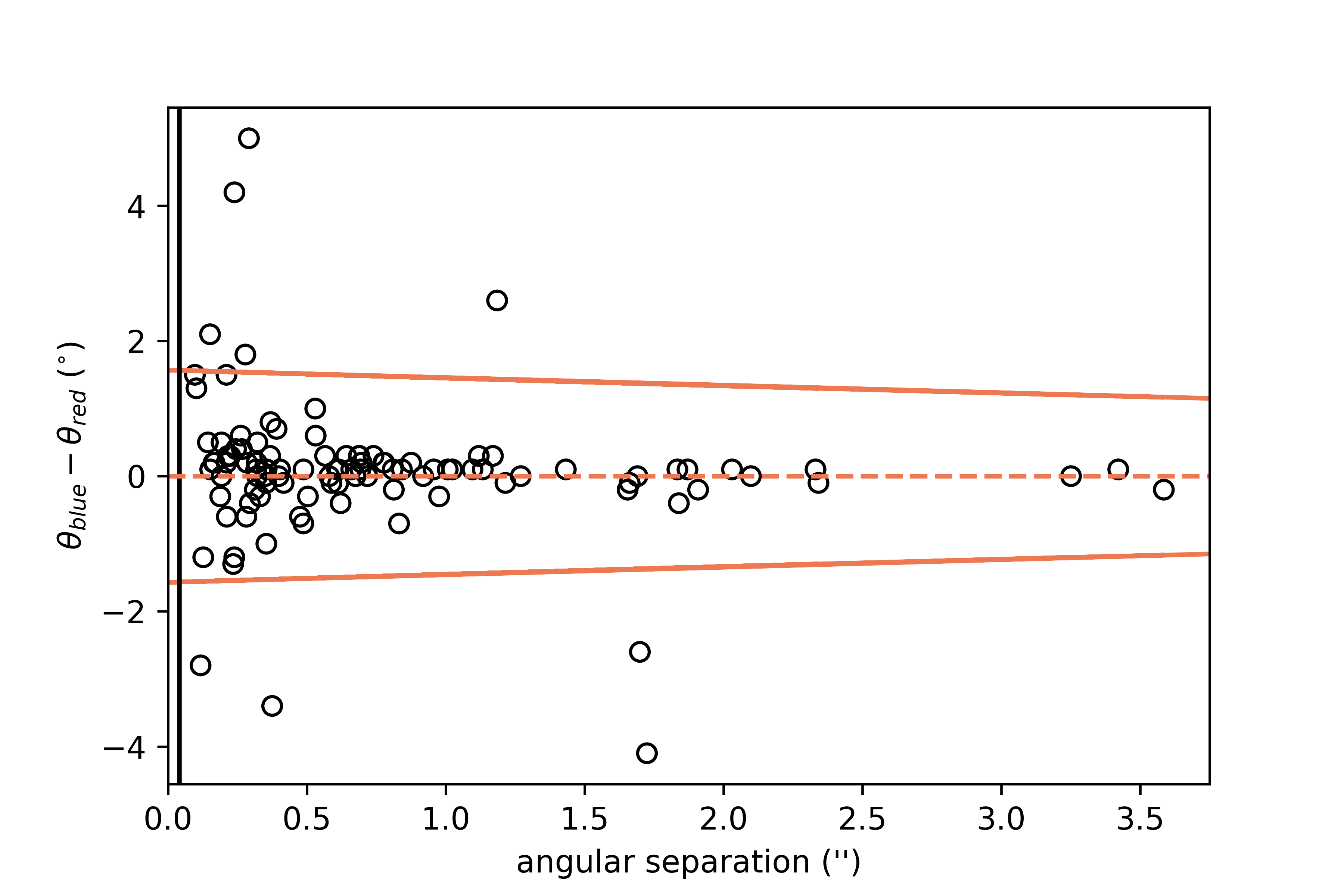}
    \includegraphics[width=0.49\textwidth]{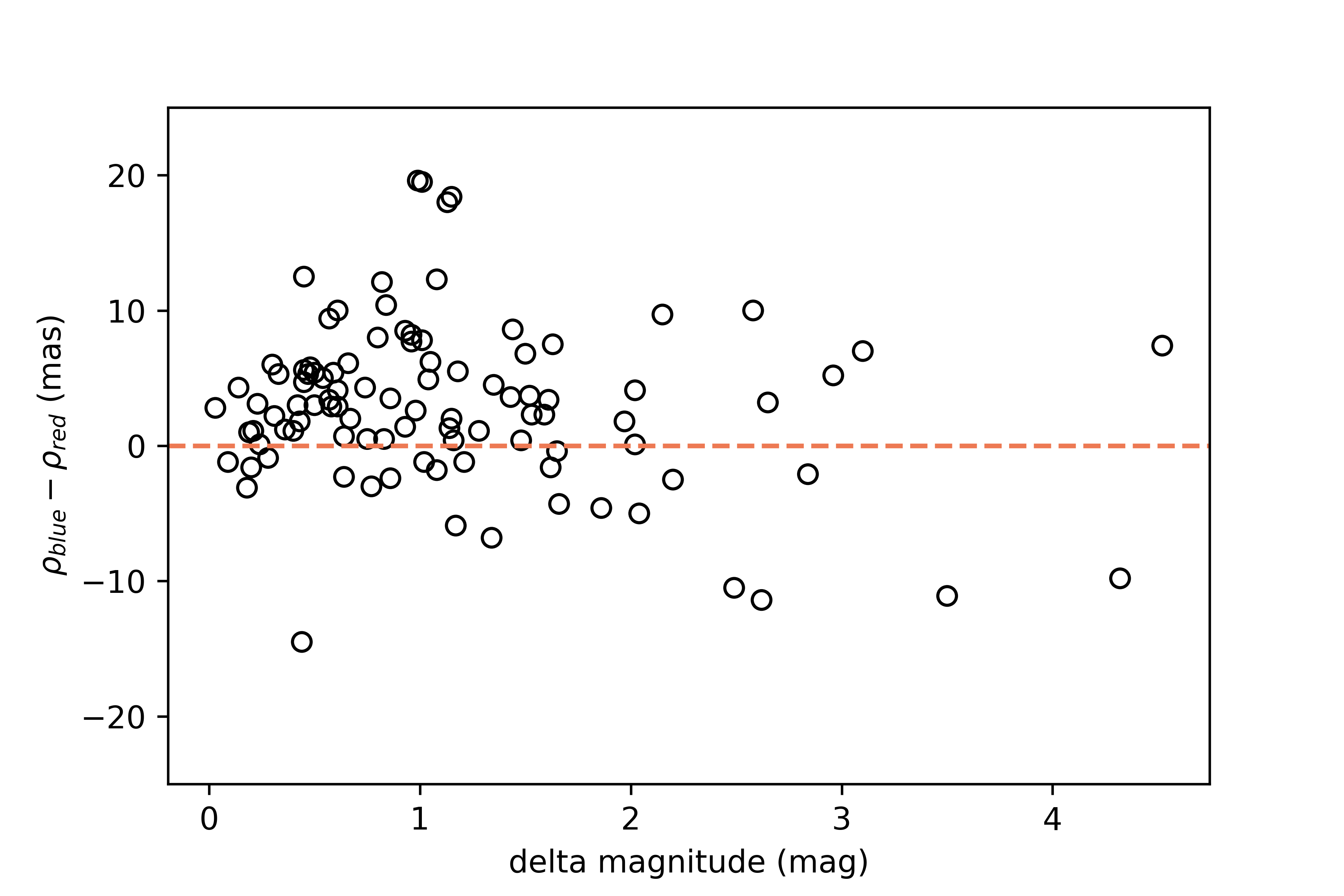}
    \includegraphics[width=0.49\textwidth]{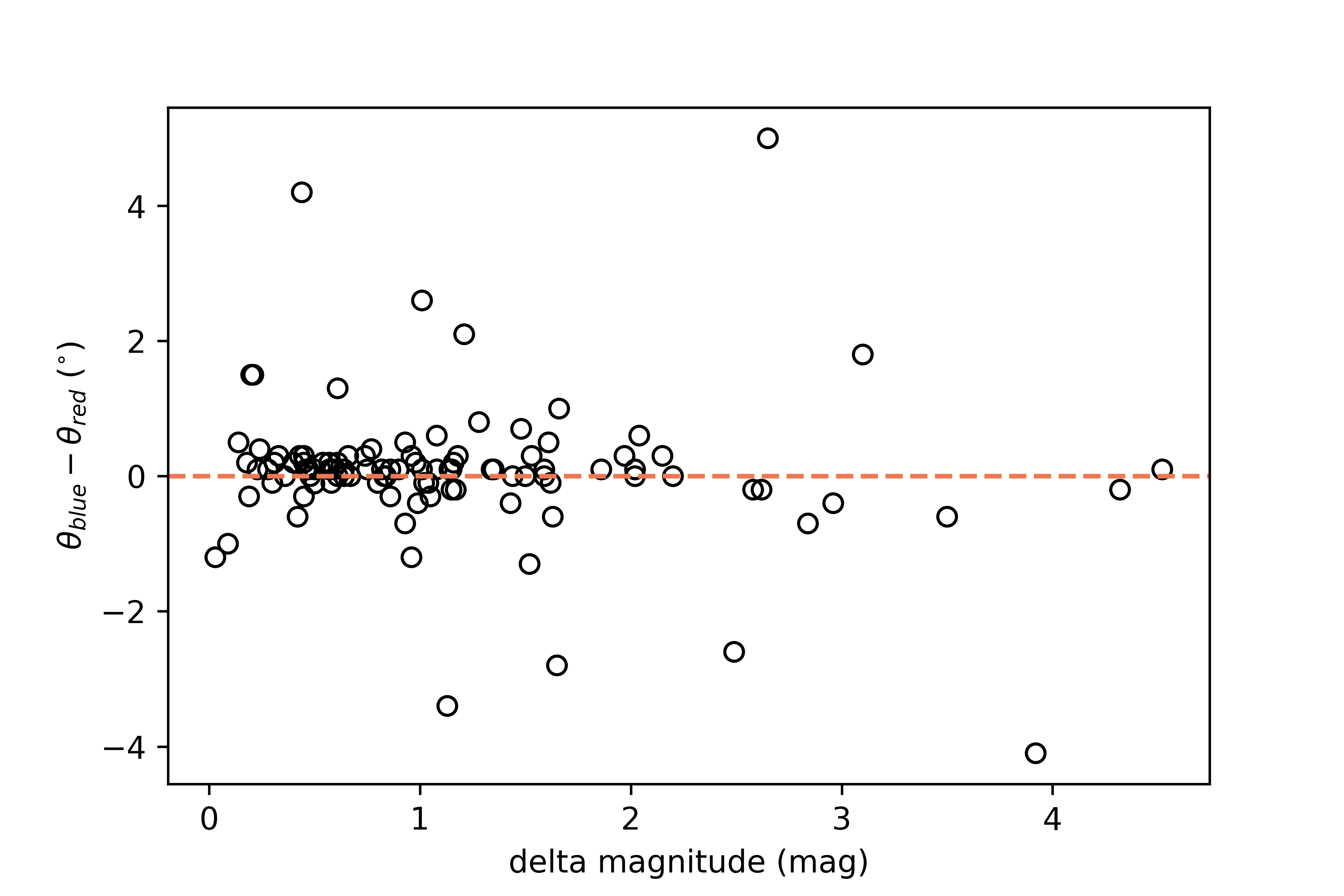}
    \caption{\textit{Top:} Differences in angular separation and position angle as a function of angular separation. In both plots, a dashed orange line at a difference of zero is shown to guide the eye, and the solid orange lines indicate the $\pm1\sigma$ in estimated internal repeatability of individual measures as a function of angular separation. For differences in separation, this value is the standard deviation of the measures, and for position angle, this value is proportional to arctan$\left(\frac{\delta\rho}{\rho}\right)$. The solid black line indicates the formal diffraction limit of the LDT. \textit{Bottom:} Differences in angular separation and position angle as a function of delta magnitude. A dashed orange line at a difference of zero is shown to guide the eye.}
    \label{fig:astrometric_precision}  
\end{figure*}

In Figure \ref{fig:astrometric_precision}, we investigate whether there are any trends in the difference between the observed properties in each channel as a function of the median of the angular separations obtained in the two channels. We find no trend in the angular separation residuals. In contrast, the position angle residuals increase as separation decreases, as the same positional uncertainty subtends a larger angle at small separations. If we assume that the positional uncertainty is the same as the orthogonal direction to the separation compared with the direction of separation itself, then we would expect the scatter in the position angle to vary as 

\begin{equation}
    \delta \theta = \text{arctan} \left(\frac{\delta \rho}{\rho}\right) = \text{arctan} \left(\frac{8.4\text{ mas}}{\rho}\right)
\end{equation}

\noindent where $\delta \theta$ is the uncertainty in the position angle difference and $\delta \rho$ is the uncertainty in the angular separation difference. We do find that the scatter in position angle as a function of angular separation is consistent with these values.

Figure \ref{fig:astrometric_precision} also shows the angular separation and position angle differences as a function of the median of the delta magnitudes obtained in the two channels; however, we do not see a trend in either of these plots. We conclude that there are no identifiable sources of systematic error in the data set.

\subsection{Photometric Precision} \label{subsec:photometric_precision}

In order to assess our photometric precision, we compare the magnitude differences derived from our speckle images to the magnitude differences from Gaia DR3. The Gaia $G$ and $R_p$ filters are reasonably similar to the 562 and 692 nm filters used throughout this survey. However, many of the companions we observed were only detected at 832 or 880 nm. Additionally, as discussed in Section \ref{sec:gaia_astrometry}, many stellar multiples with angular separations less than $\sim1\arcsec$ -- the vast majority of our detected multiples -- are unresolved in Gaia, and often do not have astrometric or photometric solutions. Taking these limitations into account, we are left with 51 observations to use for our photometric analysis.

\begin{figure*}
    \centering
    \includegraphics[width=0.49\textwidth]{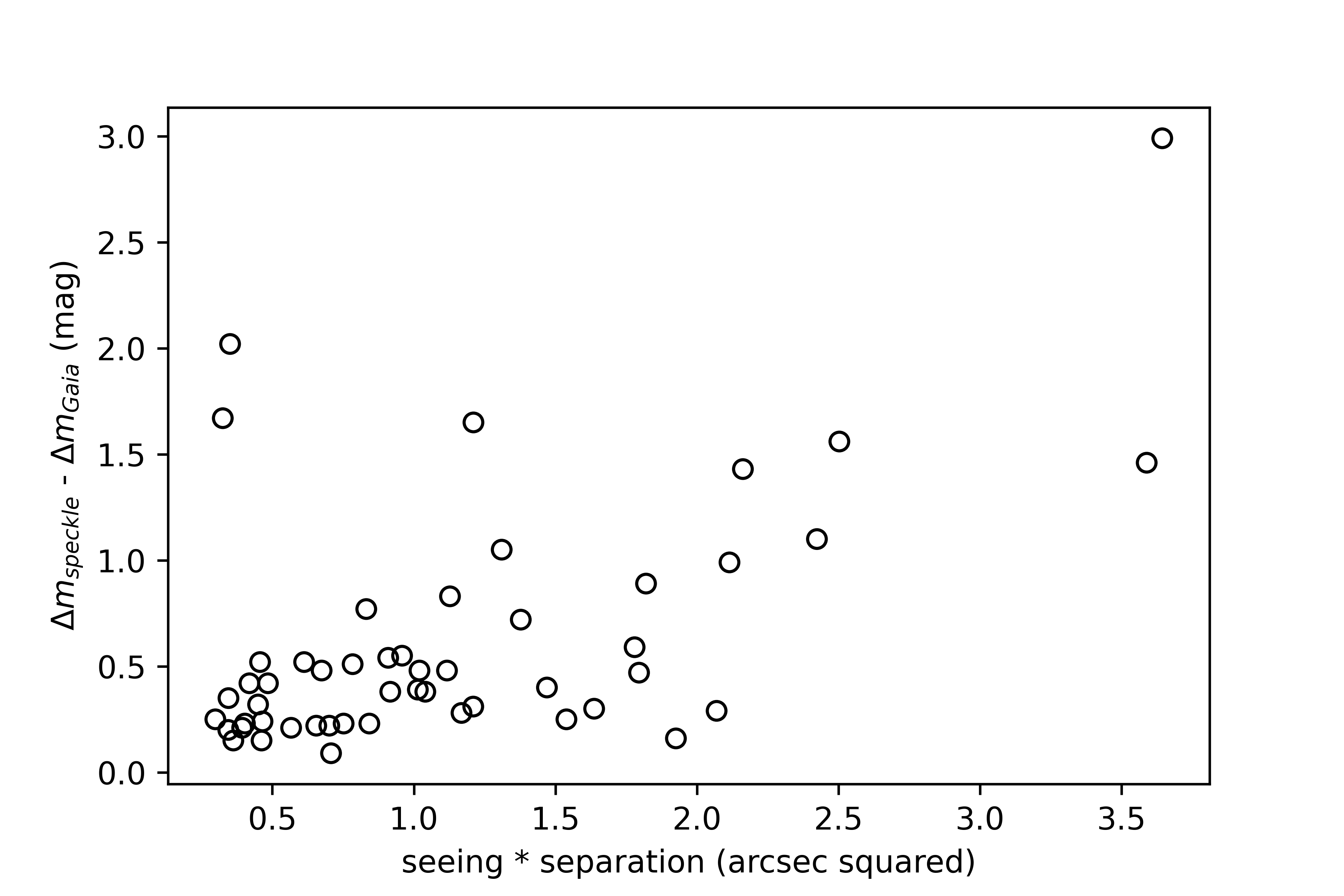}
    \includegraphics[width=0.49\textwidth]{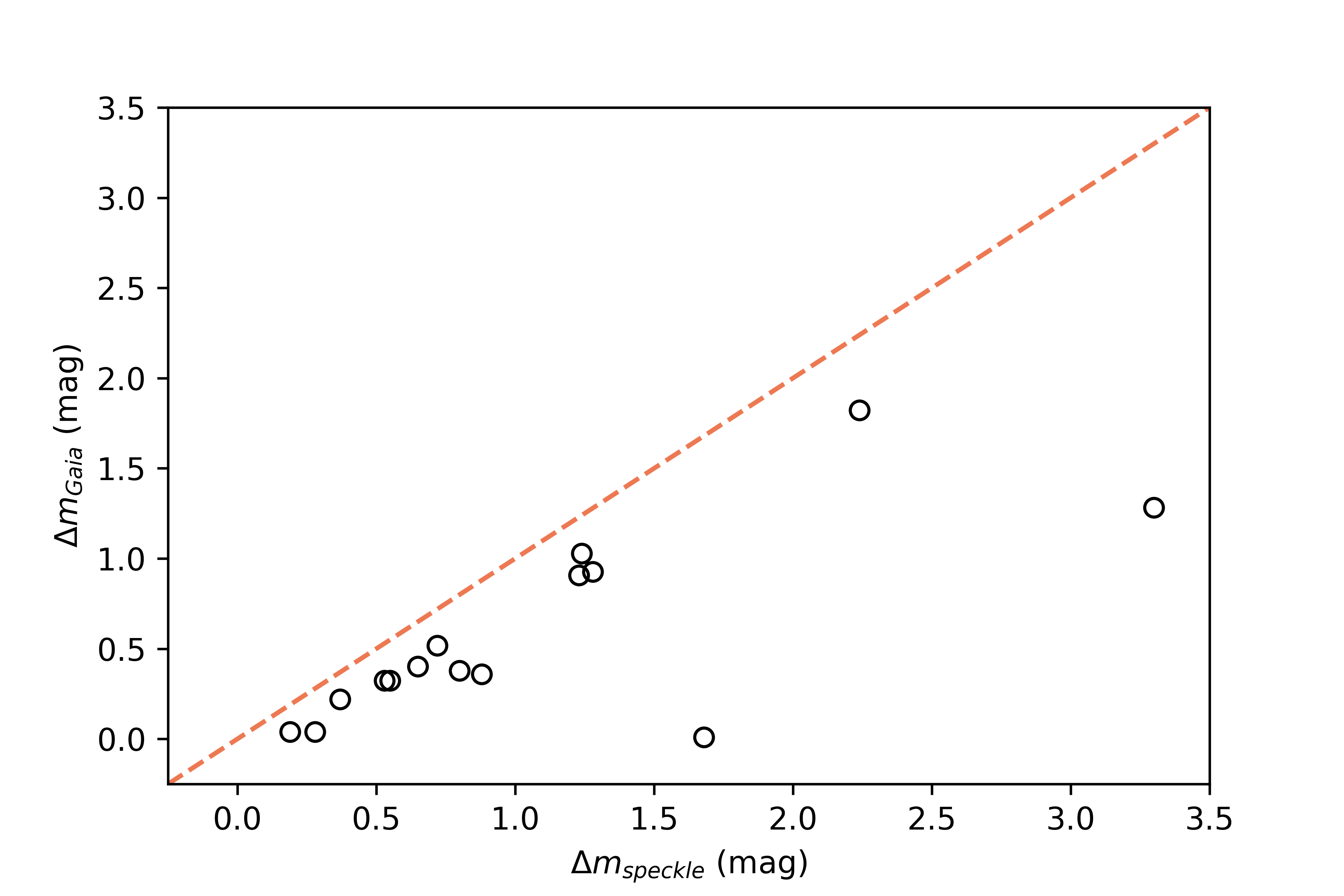}
    \caption{\textit{Left:} Difference between speckle magnitude difference and Gaia magnitude difference as function of seeing times separation. This difference between the speckle and Gaia magnitude differences should be $\sim0$ for small values of seeing times separation, and should increase as seeing times separation increases. This is the trend we see in our data. \textit{Right:} Gaia magnitude difference versus speckle magnitude difference for the observations where seeing times separation is $<0.6$ arcsec$^2$. The correlation is fairly good, although it does appear that there is a systematic offset from an equivalent relation (dashed orange line).}
    \label{fig:photometric_precision}
\end{figure*}

We show a comparison between the Gaia and speckle magnitude differences in Figure \ref{fig:photometric_precision}. We first plot the difference between the speckle magnitude difference and the Gaia magnitude difference as a function of seeing times separation, which can be used as a way to judge the isoplanicity of the observations and thus the reliability of the photometry. The larger this quantity is, the less the speckle patterns of the primary and secondary resemble one another, causing the secondary to appear fainter than it really is. The difference between the speckle and Gaia magnitude differences should be $\sim0$ for small values of seeing times separation, and should increase as seeing times separation increases. This is the trend we see in our data.

We also plot Gaia magnitude difference versus speckle magnitude difference for the 15 observations where seeing times separation is $<0.6$ arcsec$^2$. These observations should be relatively unaffected by non-isoplanicity; the speckle magnitude differences should therefore correlate well with the Gaia magnitude difference values. The correlation is fairly good, although it does appear that there is a systematic offset from an equivalent relation. The residuals from these magnitude differences have an average value of 0.25 mag, with a standard deviation of 0.54 mag. We divide by $\sqrt2$, so the average uncertainty in magnitude difference is 0.38.

\subsection{Filtered vs. Filterless Observations} \label{subsec:filterless_observations}

During our April 2017 observing run, the narrow-band speckle filters were not installed within DSSI, so the beam was simply split at $\lambda\sim700$ nm. When observing without narrow-band filters, more photons reach the detectors, resulting in additional sensitivity to faint stellar companions. However, there is also less contrast in the speckles, and significant chromatic effects arise at high airmass. Therefore, for all other observing runs, the narrow-band speckle filters were used within DSSI.

Now that all observing runs are complete, we were able to investigate whether there were any systematic effects in our data either from ambient conditions or target brightness during both the filtered and the filterless observing runs. Figure \ref{fig:median_contrast_curves} shows an increase in our limiting magnitude of $\sim0.5$ magnitudes when the speckle filters are not used. If there are no systematic effects in the filterless data, then these results indicate that filterless observations may provide better detection performance than filtered observations.

We find that the ambient conditions were comparable between the filterless observing runs and all other observing runs, although the seeing was slightly better during the filterless observing run, with a median value of 0.79 ($\sigma=0.3$) as compared to a median value of 0.90 ($\sigma=0.4$) for the filtered observing runs.

We also find that the target brightnesses were comparable between the filterless and filtered observing runs. If anything, the targets observed during the filterless observing run were a bit fainter, with a median $G$ magnitude of 12.1 ($\sigma=1.9$). The targets observed during the filtered observing runs had a median $G$ magnitude of 11.8 ($\sigma=1.6$).

Overall, 27 systems were observed both with and without the speckle filters installed. As there do not appear to be any major systematic effects that disfavor the filtered observations, we conclude that the filterless observations may provide better detection performance than filtered observations in future observing runs.

\section{Discussion} \label{sec:discussion}

In this section we evaluate the effect of stellar multiplicity on Gaia astrometry and assess any potentially-missed companions.

\newpage

\subsection{Evaluating the Effect of Stellar Multiplicity on Gaia Astrometry} \label{sec:gaia_astrometry}

The long-term astrometric monitoring from Gaia \citep{Gaia2016A&A...595A...1G, Gaia2018AA...616A...1G, Gaia2021A&A...649A...1G, Gaia2023AA...674A...1G} has proven invaluable for diverse science cases such as stellar variability, the structure of the Solar Neighborhood and the Milky Way as a whole, and in the future, even exoplanet detection. However, in the currently-available Gaia data releases, the relative motion of close-in binary systems disrupts the ability of Gaia to provide astrometric solutions for the components. This will not be the case for future Gaia data releases, as the solutions included in these data releases will use astrometric data collected over longer time baselines, and perhaps improved analysis. In the meantime, we attempt to quantify the extent to which close-in multiples affect Gaia astrometry, and show the utility of using high-resolution speckle observations to supplement and complement Gaia data.

\begin{figure*}
    \centering
    \includegraphics[width=\textwidth]{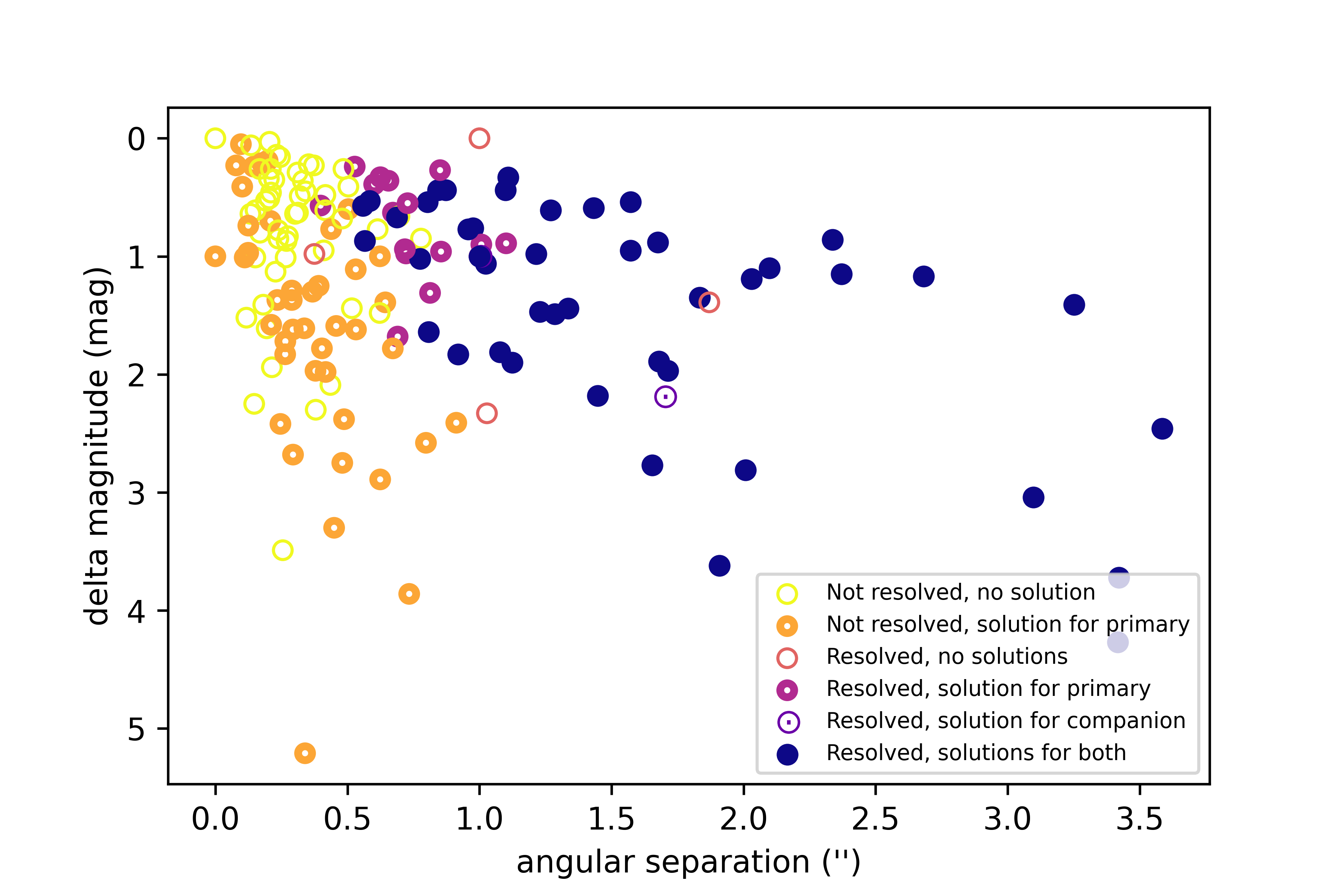}
    \caption{Scatter plot of median I-band delta magnitude versus median angular separation for the \observednum{} POKEMON targets with companions detected throughout the POKEMON survey. The different points represent whether the multiple was resolved in Gaia DR3, and whether there was an astrometric solution for the primary and for the companion. The majority (\percentresolved{}) of the companions we detected in our speckle images were \textit{not} resolved in Gaia, demonstrating the need for high-resolution imaging in addition to long-term astrometric monitoring.}
    \label{fig:delta_mag_vs_angular_sep}
\end{figure*}

In Figure \ref{fig:delta_mag_vs_angular_sep}, we plot the \observednum{} POKEMON targets with companions detected throughout the POKEMON survey. We have evaluated whether these companions were resolved in Gaia DR3, and whether there was an astrometric solution for the primary and for the companion. We find that the majority (\percentresolved{}) of the companions we detected in our speckle images were \textit{not} resolved in Gaia, demonstrating the need for high-resolution imaging in addition to long-term astrometric monitoring. In general, the companions that were not resolved by Gaia were within $1\arcsec$ and had large delta magnitudes.

However, multiple Gaia parameters are known to indicate the presence of a stellar companion. First, the Gaia re-normalized unit weight error (RUWE) acts like a reduced chi-squared, where large values can indicate a poor model fit to the astrometry, assuming that the star is single. Single sources typically have RUWE values of $\sim1$, while sources with RUWE values $>1.4$ are likely non-single or otherwise extended \citep{Ziegler2020AJ....159...19Z, Gaia2021A&A...649A...1G}. Following \citet{Vrijmoet2020AJ....160..215V}, which surveyed M dwarfs specifically, we use RUWE $>2$ to distinguish single and (potentially) non-single sources. Additionally, the ipd\_frac\_multi\_peak (IPDFMP) value provides the fraction of windows -- as a percentage from 0 to 100 -- for which the image parameter determination (IPD) algorithm has identified a double peak, meaning that the detection may be a visually resolved double star. Following \citet{Tokovinin2023AJ....165..180T}, we use IPDFMP $>2$ to distinguish single and (potentially) non-single sources.

\begin{figure*}
    \centering
    \includegraphics[width=0.49\textwidth]{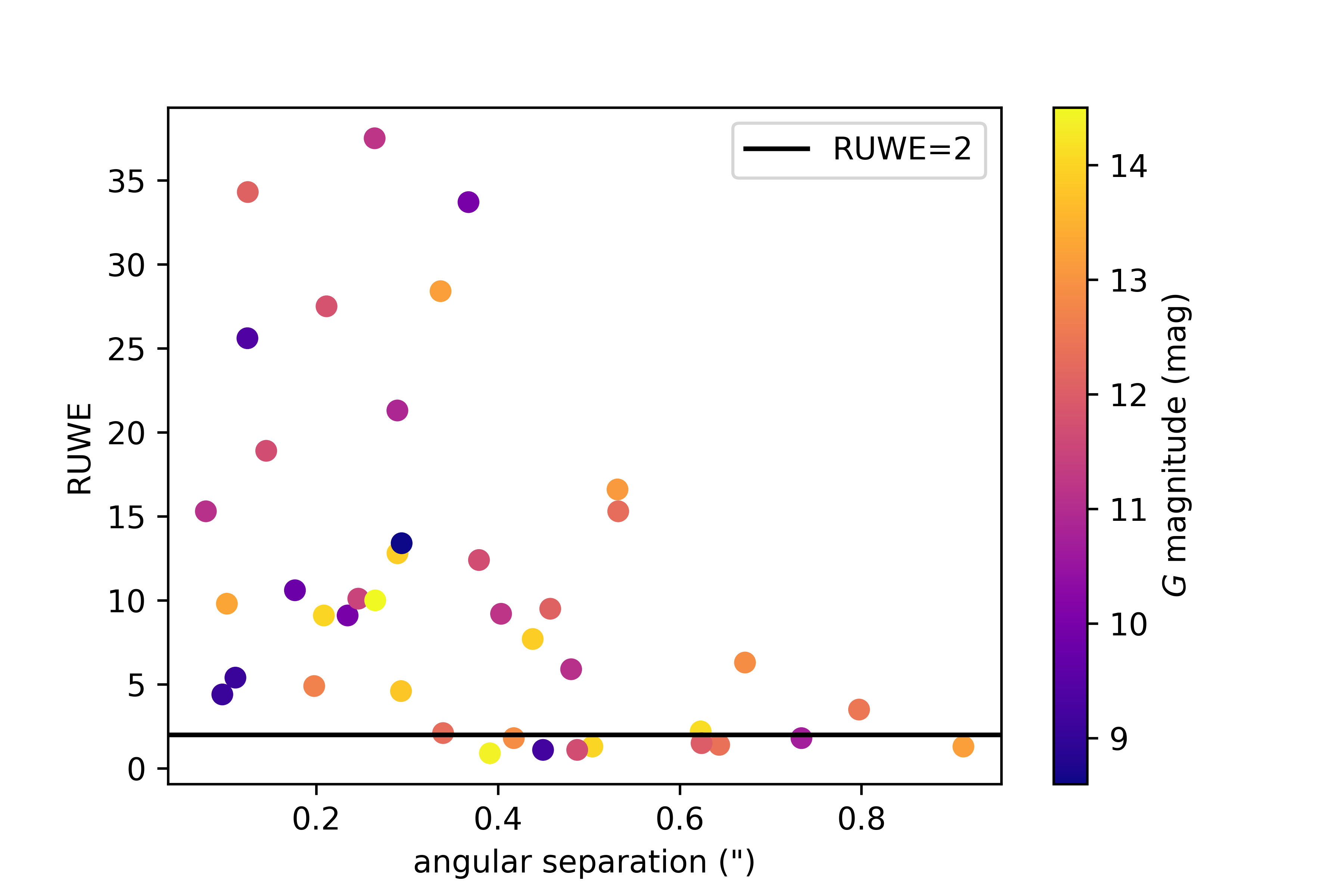}
    \includegraphics[width=0.49\textwidth]{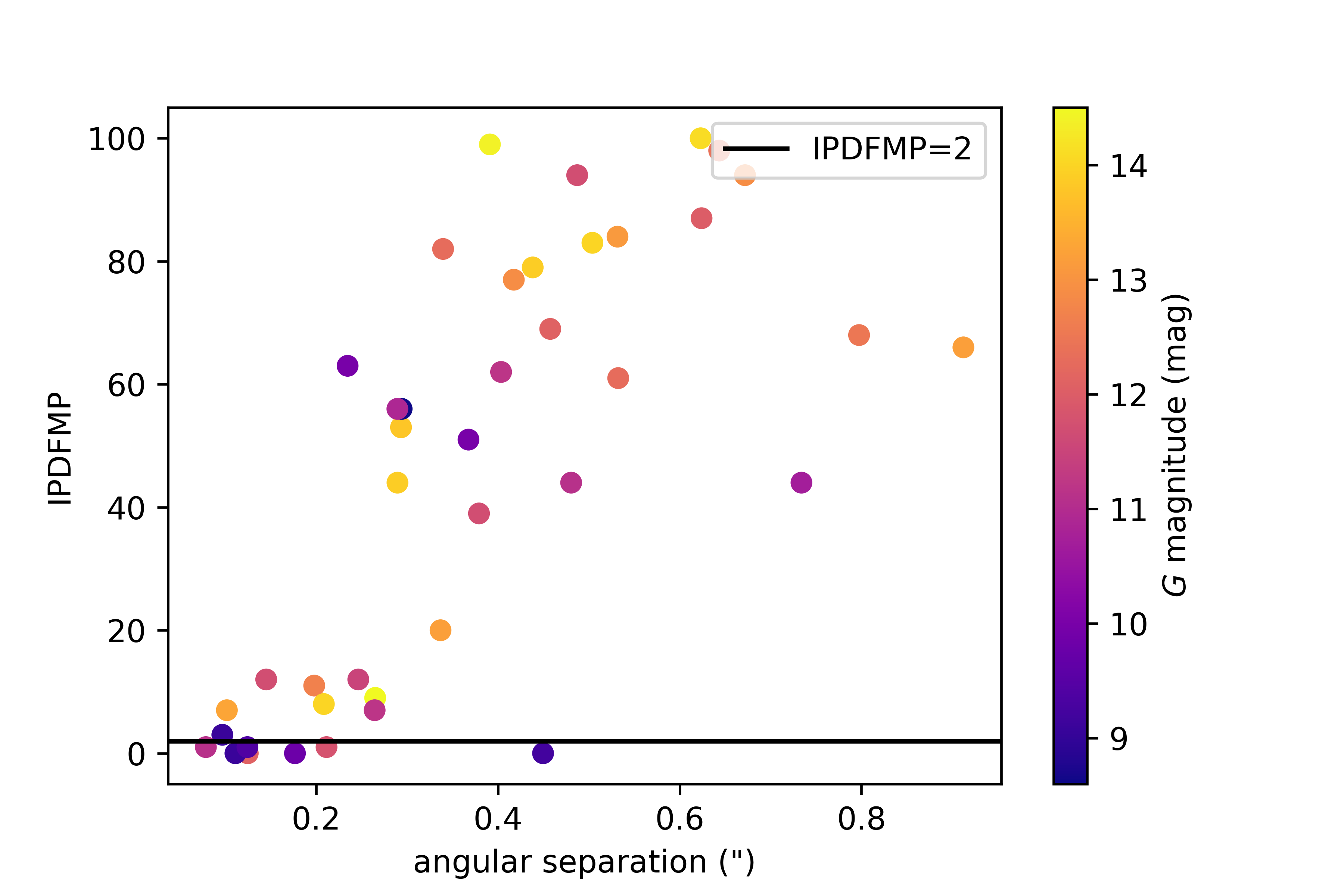}
    \caption{Scatter plot of Gaia RUWE value (left) and IPDFMP value (right) versus angular separation for the 41 POKEMON multiples that were not resolved in Gaia DR3, but for which the primary has a Gaia DR3 astrometric solution. The points are color-coded by Gaia $G$ magnitude, and the black line indicates a RUWE value of 2. All of the unresolved multiples are within $1\arcsec$. While the RUWE does not reveal the presence of an unseen companion in every case, we do find that the majority (78.0\% and 82.9\%, respectively) of the RUWE and IPDFMP values are greater than 2, and that the median RUWE and IPDFMP values are 9.1 and 51, respectively. These results indicate that the RUWE and IPDFMP value can be useful identifiers of unseen stellar companions. In particular, the RUWE value is useful for close-in multiples, and the IPDFMP value is useful for wider multiples. However, these results also indicate that high-resolution imaging -- in addition to long-term astrometric monitoring from Gaia -- is critical for both resolving stellar multiples and determining the properties of any unseen companions.}
    \label{fig:ruwe_vs_angular_sep}
\end{figure*}

In order to understand how often a high RUWE or IPDFMP value indicates the presence of an unseen companion, we plot the RUWE and IPDFMP values as a function of angular separation for the 41 POKEMON multiples that were not resolved in Gaia DR3, but with a Gaia DR3 astrometic solution in Figure \ref{fig:ruwe_vs_angular_sep} (48 unresolved POKEMON multiples did not have Gaia DR3 astrometric solutions). We find that all of the unresolved companions whose primaries have Gaia DR3 astrometric solutions are within $1\arcsec$. While the RUWE and IPDFMP values do not reveal the presence of an unseen companion in every case, we do find that the majority (78.0\%) of the RUWE values are greater than 2, and that the median RUWE value is 9.1. We find that 82.9\% of the IPDFMP values are greater than 2, and that the median IPDFMP value is 51. These results indicate that even though Gaia astrometry can become unreliable for close-in stellar multiples, the RUWE and IPDFMP values can still be a useful identifier of potential unseen stellar companions. However, these results also indicate that high-resolution imaging -- in addition to long-term astrometric monitoring from Gaia -- is critical for both resolving stellar multiples and determining the properties of any unseen companions.

We also investigated whether any of the stars we have deemed single have elevated RUWE and IPDFMP values, indicating that they may have an as-yet undetected stellar companion. We find that 12 of the ``single'' stars in our sample have both RUWE and IPDFMP values over 2; we list these targets Table \ref{table:potential_multiples}. These stars are potential targets for additional high-resolution and/or high-contrast imaging, or spectroscopic follow-up.

\startlongtable
\begin{deluxetable}{ccc}
\tablecaption{Potential multiples
\label{table:potential_multiples}}
\tablehead{\colhead{2MASS ID} & \colhead{RUWE} & \colhead{IPDFMP}}
\startdata
06043887+0741545 & 3.1 & 63 \\
06170531+8353354 & 17.0 & 76 \\
08115757+0846220 & 2.1 & 4 \\
08255285+6902016 & 4.6 & 53 \\
08505062+5253462 & 5.4 & 67 \\
08561768-2326574 & 2.9 & 98 \\
11065691-1244024 & 8.9 & 3 \\
12394672+0410471 & 3.2 & 21 \\
13525004+6537197 & 2.4 & 73 \\
14442602-1800065 & 14.4 & 82 \\
20125995+0112584 & 7.9 & 4 \\
23373831-1250277 & 6.6 & 12 \\
\enddata
\end{deluxetable}

\newpage

\subsection{Assessing Potentially-Missed Companions} \label{subsec:missed_companions}

In order to identify the population of stellar companions that were not detected in our speckle images -- such as those indicated by a lack of a Gaia astrometric solution or a high RUWE value -- we used existing code that was originally developed for Palomar and Keck adaptive optics observations \citep{LundCiardi2020AAS...23524906L} and adapted it to the specific needs of the POKEMON survey and our speckle observations. The purpose of these simulations is to estimate the fraction of stellar companions that would be detectable within our speckle images, assuming that the star has a companion.

\begin{figure*}
    \centering
    \includegraphics[width=\textwidth]{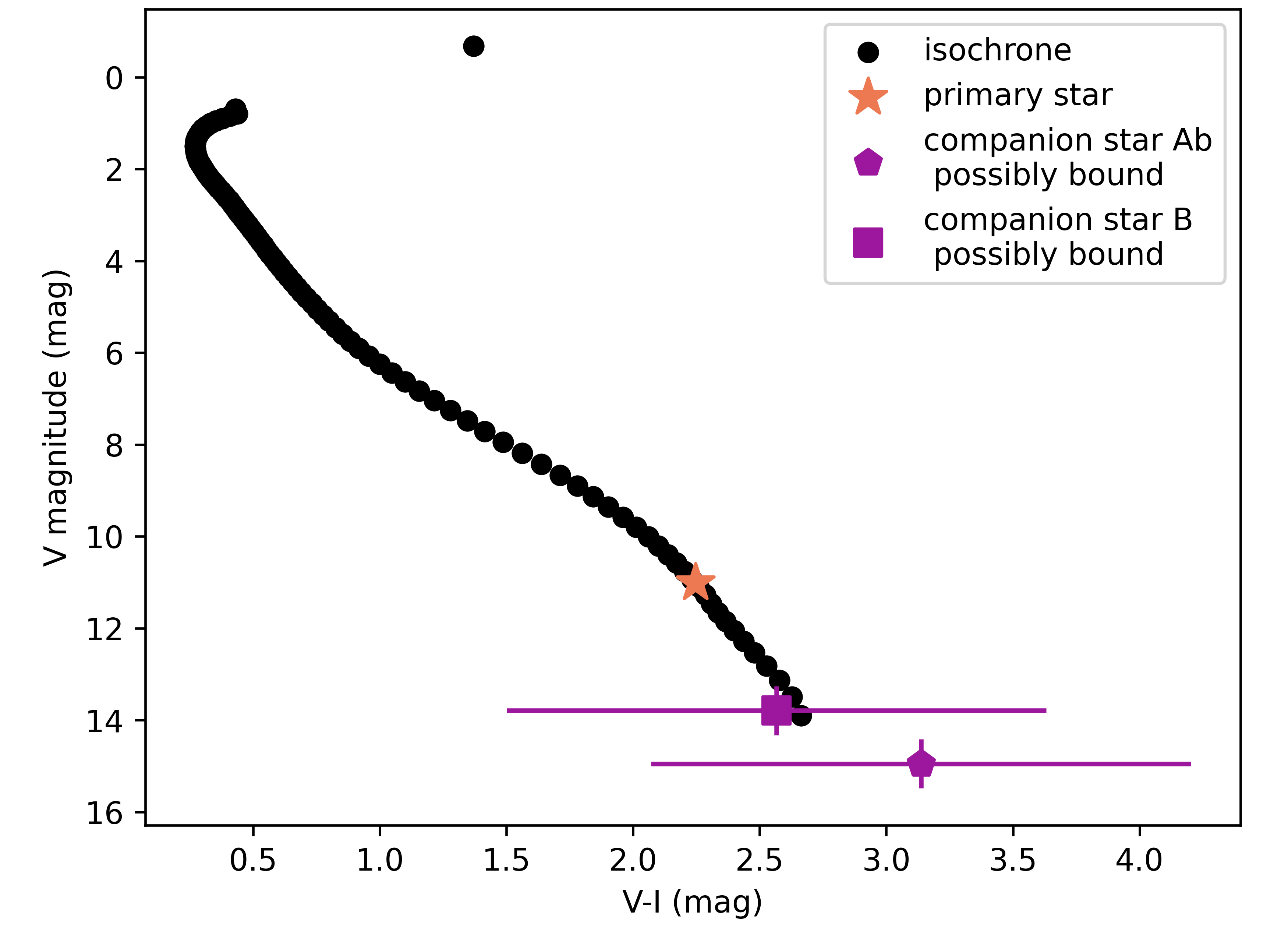}
    \caption{Example of the isochrone fitting carried out to estimate the stellar mass range that may have been missed by our high-resolution imaging. This color-magnitude diagram corresponds to 2MASS J00322970+6714080, and the plotted isochrone corresponds to an age of 0.9 Gyr. The distance to this object is ten pc, so here the apparent and absolute magnitudes are identical. However, it should be noted that if the companions are unbound stars at an arbitrary distance, then they may be far from the isochrone that fits the primary in absolute magnitude space. Nonetheless, the fact that the companions are self-consistently compatible with having the same distance as the primary still has informational value. While color uncertainties of the secondary and tertiary components in this system are still large at this point, the placement of these objects on the color-magnitude diagram is consistent with a gravitationally-bound triple system.}
    \label{fig:color-mag_diagram}
\end{figure*}

The code works by first identifying the population of stellar companions that could orbit each star, and then uses the derived contrast curves to evaluate the sensitivity of each observation to these stellar companions. Specifically, the code identifies the populations of stellar companions by matching the star to a best-fit stellar isochrone from the Dartmouth isochrones \citep{Dotter2008ApJS..178...89D}. The primary goal of the isochrone fitting is to estimate the stellar mass range that may have been missed by the high-resolution imaging. Given that the majority of the POKEMON targets do not have age estimates, the age of the primary star is not set \textit{a priori}, and the isochrone fitting is done across all of the possible Dartmouth models (0.25 – 15 Gyr). The fitting yielded stellar ages that are consistent with the M-dwarf Main Sequence, but that were not actually used to age the stars. Given the long lifetime of the M dwarfs, uncertainties in the isochrone fitting due to age are dwarfed by the general uncertainties in the photometric fitting, so the isochrone fitting yields reasonable estimates of the mass range that remained undetected. See Figure \ref{fig:color-mag_diagram} for additional discussion of the isochrone fitting.

The code then independently draws from the mass ratio and orbital period distributions for M dwarfs from the \citet{DucheneKraus2013ARA&A..51..269D} review paper on stellar multiplicity. The code makes the assumption that if the companion is outside the field-of-view of the instrument, then it would be revealed by other methods. We also account for any known companions (when their properties are available), as they could influence the population of stellar companions that could orbit their host star.

\begin{figure*}
    \centering
    \includegraphics[width=\textwidth]{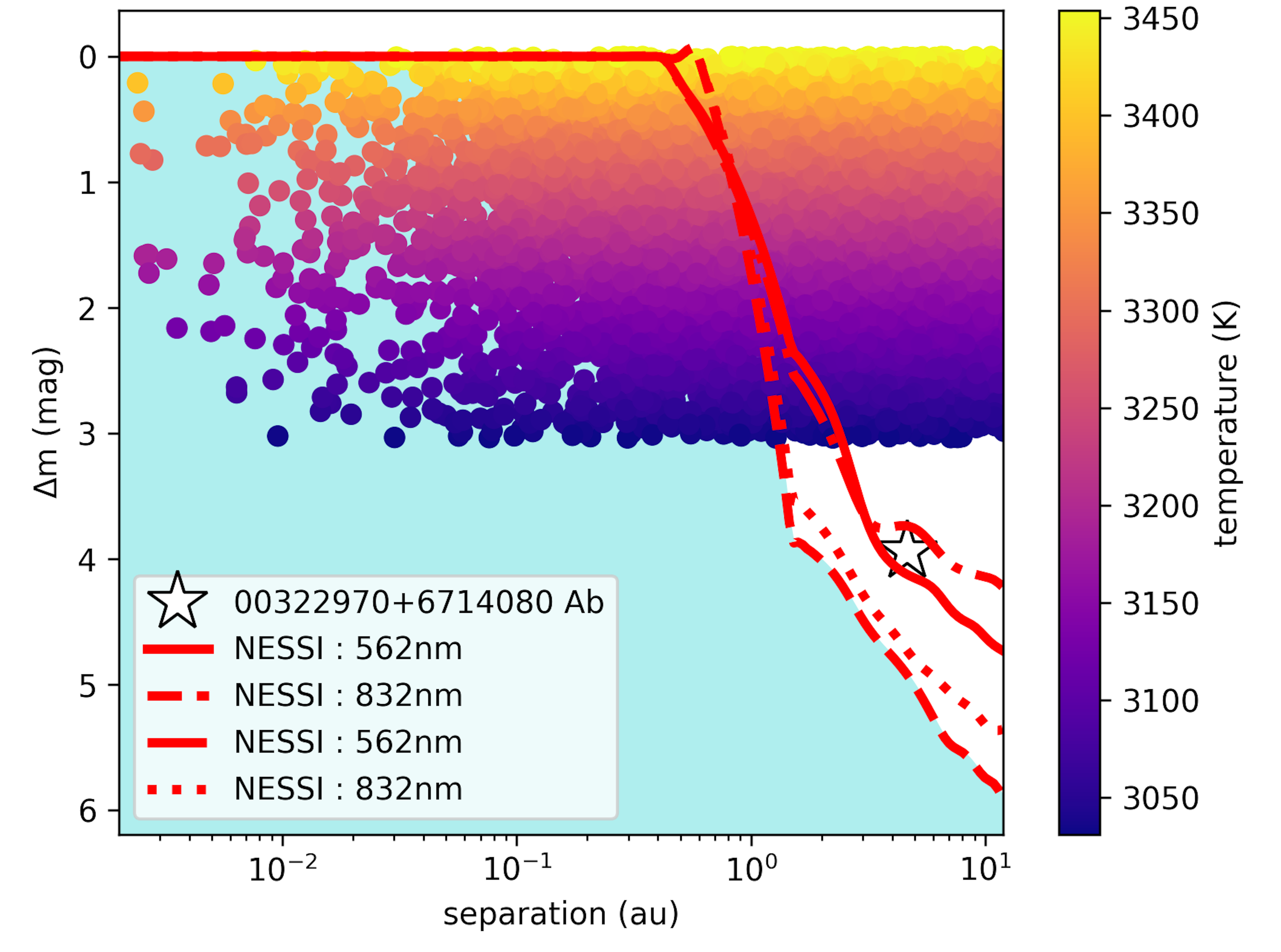}
    \caption{Simulated companions and derived contrast curves for the triple system 2MASS J00322970+6714080. For this particular star, 76.5\% of the simulated stellar companions would be detectable within our speckle images, and we did indeed detect both companions. The inner companion to this star is shown here near the limits of the contrast curves.}
    \label{fig:simulated_companions}
\end{figure*}

An example of the simulated companions is shown in Figure \ref{fig:simulated_companions}. The separations for the simulated companions are calculated assuming a circular orbit, and using the period and stellar masses to compute a semi-major axis. The companion is then given a random position on that orbit. This simulation is performed 10,000 times for each star. These simulations allow us to determine the fraction of stellar companions that would be detectable within our speckle images.

\begin{figure*}
    \centering
    \includegraphics[width=\textwidth]{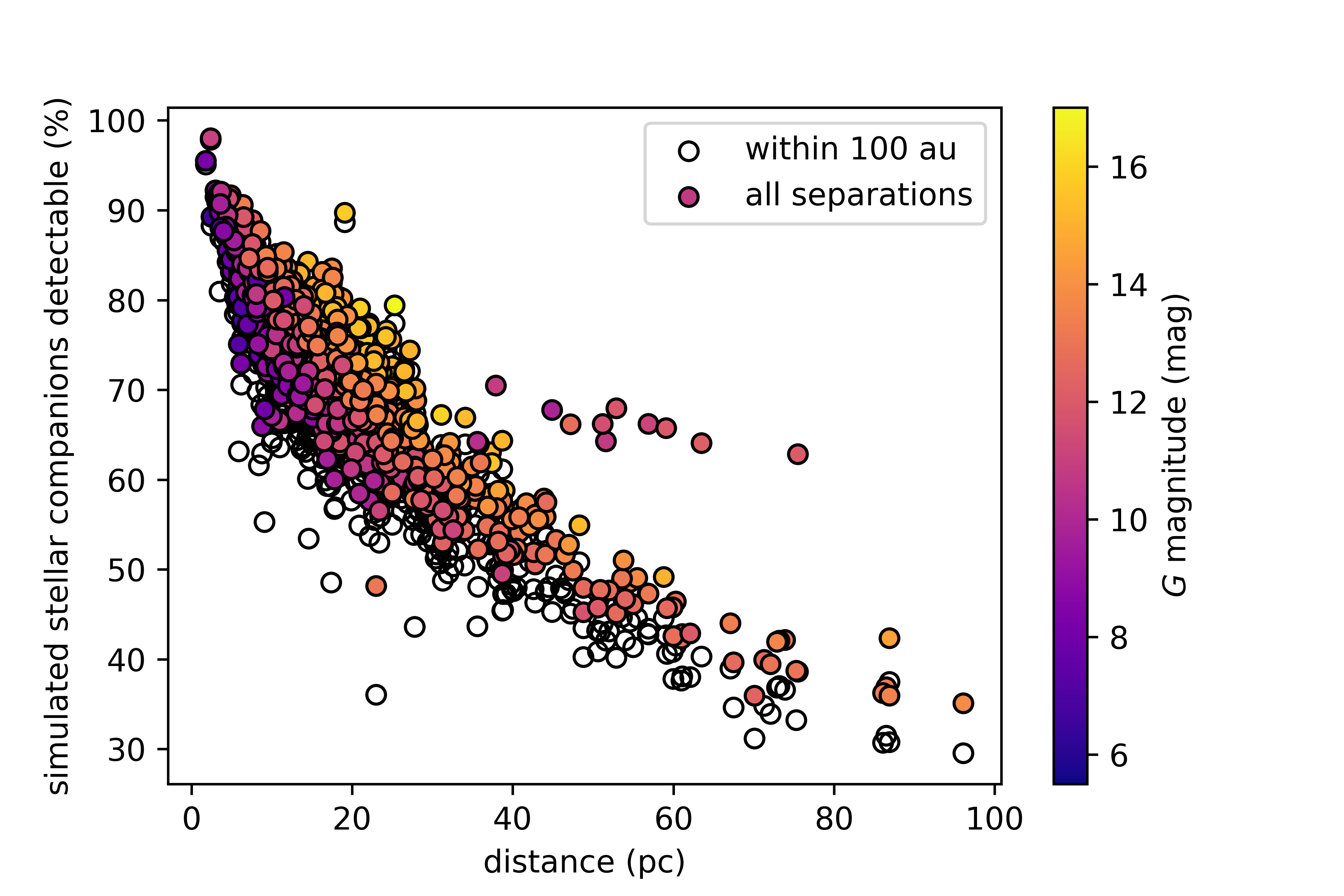}
    \caption{Percent of stellar companions detectable within our speckle images versus distance. Circles color-coded by $G$ magnitude represent the detectability of simulated companions at all separations. Open black circles represent the detectability of simulated companions specifically within 100 au. We find that the majority of potential stellar companions would be detectable, particularly for the nearby targets, with a median value of \percentdetectable{}. Specifically within 100 au, we find that \percentdetectablehundred{} of simulated companions are recovered.}
    \label{fig:percent_detectable}
\end{figure*}

We note that 70 of the \pokemonnum{} POKEMON targets were unable to be included in the \citet{LundCiardi2020AAS...23524906L} analysis, either because their mass was too low to be fit with a Dartmouth isochrone, because their mass was too low to simulate a meaningful number of potential stellar companions, or because we were unable to determine a mass or distance for the target. Nonetheless, we were able to carry out this analysis on 93.8\% of the targets in the POKEMON sample. For these targets, we find that the majority of potential stellar companions would be detectable by our speckle observations, particularly for the nearby targets (Figure \ref{fig:percent_detectable}); we find a median value of \percentdetectable{}. Specifically within 100 au, we find that \percentdetectablehundred{} of simulated companions are recovered. In general, the simulated companions that were not detected by our technique are either very close-in, or are much fainter than their primary stars.

We note that neither the Dartmouth isochrones nor the \citet{DucheneKraus2013ARA&A..51..269D} distributions are perfect for simulating companions to the POKEMON sample, as the Dartmouth isochrones only go through M6V, and the \citet{DucheneKraus2013ARA&A..51..269D} have a lower limit for the primary star at $0.1M_{\odot}$. In addition, as described previously, we consider our speckle photometry approximate. Nonetheless, we deem these resources the most appropriate to assess the completeness of our observations based on what is currently available in the literature. However, these biases do point to the necessity of a comprehensive M-dwarf multiplicity survey that examines the distributions of stellar companions to M dwarfs at all mass ratios and orbital periods. Ongoing work that will be presented in a future paper in the series will provide these updated mass ratio and orbital period distributions; this work is discussed further in Section \ref{sec:conclusions}.

\section{Conclusions and Future Work} \label{sec:conclusions}

We have carried out the POKEMON speckle survey of nearby M dwarfs using the DSSI and NESSI instruments on the 4.3-meter LDT and the 3.5-meter WIYN telescope, respectively. Using these instruments, we imaged \pokemonnum{} targets, and have revealed companions to \observednum{} of them.

We investigate the effect of this stellar multiplicity on Gaia astrometry, and find that the majority (\percentresolved{}) of the companions we detect in our speckle images were \textit{not} resolved in Gaia, demonstrating the need for high-resolution imaging in addition to long-term astrometric monitoring. We do, however, argue that the Gaia RUWE and IPDFMP values can be a useful tool for assessing the likelihood of an unseen stellar companion, and we note 14 potentially-multiple systems identified via these metrics.

We also find that the majority (\percentdetectable{}) of simulated stellar companions would be detectable by our speckle observations, particularly for the nearby targets. Specifically within 100 au, we find that \percentdetectablehundred{} of simulated companions are recovered.

We are continuing to follow up the POKEMON targets with the Quad-camera Wavefront-sensing Six-channel Speckle Interferometer \citep{Clark2020SPIE11446E..2AC}. Additionally, upcoming POKEMON papers will present the stellar multiplicity rate of M dwarfs within 15 pc (Clark et al. submitted), as well as the stellar multiplicity rate of M dwarfs within 15 pc calculated by spectral subtype through M9V for the first time.

\clearpage


We thank the anonymous reviewer, Davy Kirkpatrick, Chris Gelino, Alex Polanski, Mark Popinchalk, and Frederick Hahne for their insightful contributions to this manuscript. We are also thankful to the army of TOs at the LDT and the WIYN Telescope for all of their assistance during our 50 nights of observing.

This research was carried out at the Jet Propulsion Laboratory, California Institute of Technology, under a contract with the National Aeronautics and Space Administration (80NM0018D0004). This research was supported by NSF Grant No. AST-1616084 awarded to GTvB and NASA Grant 18-2XRP18\_2-0007 awarded to DRC.

These results made use of the Lowell Discovery Telescope at Lowell Observatory. Lowell is a private, nonprofit institution dedicated to astrophysical research and public appreciation of astronomy and operates the LDT in partnership with Boston University, the University of Maryland, the University of Toledo, Northern Arizona University, and Yale University. Lowell Observatory sits at the base of mountains sacred to tribes throughout the region. We honor their past, present, and future generations, who have lived here for millennia and will forever call this place home.

These results are also based on observations from Kitt Peak National Observatory, the NSF's National Optical-Infrared Astronomy Research Laboratory (NOIRLab Prop. ID: 2018B-0126; PI: C. Clark), which is operated by the Association of Universities for Research in Astronomy (AURA) under a cooperative agreement with the National Science Foundation. The authors are honored to be permitted to conduct astronomical research on Iolkam Du’ag (Kitt Peak), a mountain with particular significance to the Tohono O’odham. Data presented herein were obtained at the WIYN Observatory from telescope time allocated to NN-EXPLORE through the scientific partnership of the National Aeronautics and Space Administration, the National Science Foundation, and the NSF's National Optical-Infrared Astronomy Research Laboratory. Observations in the paper made use of the NN-EXPLORE Exoplanet and Stellar Speckle Imager (NESSI). NESSI was funded by the NASA Exoplanet Exploration Program and the NASA Ames Research Center. NESSI was built at the Ames Research Center by Steve B. Howell, Nic Scott, Elliott P. Horch, and Emmett Quigley.

This work presents results from the European Space Agency (ESA) space mission Gaia
. Gaia data are being processed by the Gaia Data Processing and Analysis Consortium (DPAC). Funding for the DPAC is provided by national institutions, in particular the institutions participating in the Gaia MultiLateral Agreement (MLA). The Gaia mission website is \url{https://www.cosmos.esa.int/gaia}. The Gaia archive website is \url{https://archives.esac.esa.int/gaia}.

This work has used data products from the Two Micron All Sky Survey \citep{https://doi.org/10.26131/irsa2}, which is a joint project of the University of Massachusetts and the Infrared Processing and Analysis Center at the California Institute of Technology, funded by NASA and NSF.

This research has made use of the Exoplanet Follow-up Observation Program (ExoFOP, DOI:10.26134/ExoFOP5) website, which is operated by the California Institute of Technology, under contract with the National Aeronautics and Space Administration under the Exoplanet Exploration Program.

Information was collected from several additional large database efforts: the Simbad database and the VizieR catalogue access tool, operated at CDS, Strasbourg, France; NASA's Astrophysics Data System; and the Washington Double Star Catalog maintained at the US Naval Observatory.

© 2023. California Institute of Technology. Government sponsorship acknowledged. 


%

\vspace{5mm}
\facilities{LDT(DSSI), WIYN(NESSI)}


\software{Astropy \citep{Astropy2013}, Astroquery \citep{Astroquery2019AJ....157...98G}, IPython \citep{IPython2007}, Matplotlib \citep{Matplotlib2007}, NumPy \citep{NumPy2020}, Pandas \citep{Pandas2010}, SciPy \citep{SciPy2020}}

\clearpage





\bibliography{references}{}
\bibliographystyle{aasjournal}



\end{document}